\documentclass[10pt,aps,prc,longbibliography,nofootinbib,showkeys,superscriptaddress,twocolumn,floatfix]{revtex4-2}

\usepackage{hyperref}
\usepackage{graphicx}

\usepackage{lineno}
\usepackage{xspace}
\usepackage{comment}
\usepackage{multirow}
\usepackage{array,booktabs}

\usepackage{amsmath,amssymb}

\usepackage{rotating}
\usepackage{bm}

\usepackage[normalem]{ulem}

\graphicspath{ {./figures/} }

\usepackage{xcolor}
\definecolor{myOrange}{rgb}{1,0.5,0.}
\definecolor{myGreen}{rgb}{0.0,0.6,0.1}

\newcommand{\pp}           {pp\xspace}

\newcommand{\PbPb}         {\mbox{Pb--Pb}\xspace}

\newcommand{\RAA}          {\ensuremath{R_{\mathrm{AA}}}\xspace}

\newcommand{\TAA}          {\ensuremath{\langle{T}_{\mathrm{AA}}\rangle}\xspace}
\newcommand{\dNAA}         {\ensuremath{\mathrm{d}N_\mathrm{AA}}\xspace}
\newcommand{\dsigpp}       {\ensuremath{\mathrm{d}\sigma_\mathrm{pp}}\xspace}
\newcommand{\dpT}       {\ensuremath{\mathrm{d}{\pT}}\xspace}

\newcommand{\nineH}        {$\sqrt{s}~=~0.9$~Te\kern-.1emV\xspace}
\newcommand{\seven}        {$\sqrt{s}~=~7$~Te\kern-.1emV\xspace}
\newcommand{\twoH}         {$\sqrt{s}~=~0.2$~Te\kern-.1emV\xspace}
\newcommand{\twosevensix}  {$\sqrt{s}~=~2.76$~Te\kern-.1emV\xspace}
\newcommand{\five}         {$\sqrt{s}~=~5.02$~Te\kern-.1emV\xspace}
\newcommand{\twosevensixnn}{$\sqrt{s_{\mathrm{NN}}}~=~2.76$~Te\kern-.1emV\xspace}
\newcommand{\fivenn}       {$\sqrt{s_{\mathrm{NN}}}~=~5.02$~Te\kern-.1emV\xspace}

\newcommand{\GeVc}         {Ge\kern-.1emV/$c$\xspace}
\newcommand{\MeVc}         {Me\kern-.1emV/$c$\xspace}
\newcommand{\TeV}          {Te\kern-.1emV\xspace}
\newcommand{\GeV}          {Ge\kern-.1emV\xspace}
\newcommand{\MeV}          {Me\kern-.1emV\xspace}
\newcommand{\GeVmass}      {Ge\kern-.2emV/$c^2$\xspace}
\newcommand{\MeVmass}      {Me\kern-.2emV/$c^2$\xspace}

\newcommand{\qsqr}{\ensuremath{Q^2}}

\newcommand{\qhat}{\ensuremath{\hat{q}}\xspace}

\newcommand{\alphas}{\ensuremath{\alpha_{\text{s}}}\xspace}

\newcommand{\alphasfix}{\ensuremath{\alpha_{\text{s}}^\mathrm{fix}}\xspace}

\newcommand{\qswitch}{\ensuremath{Q_{\text{0}}}\xspace}
\newcommand{\tstart}{\ensuremath{\tau_{\text{0}}}\xspace}

\newcommand{\sqrtsNN}{\ensuremath{\sqrt{s_\mathrm{NN}}}}

\newcommand{\pT}{\ensuremath{p_\mathrm{T}}}

\newcommand{\URQMD}{\textsc{UrQMD}}
\newcommand{\PYTHIA}{\textsc{Pythia}}

\newcommand{\TRENTO}{\textsc{Trento}}

\newcommand{\STi}{\ensuremath{S_{T_l}}}
\newcommand{\STinorm}{\ensuremath{S_{T_l (\rm norm)}}}

\newcommand{\XbfTildei}{\ensuremath{\mathbf{X}_{\sim{l}}}}
\newcommand{\Xbf}{\ensuremath{\mathbf{X}}}

\newcommand{\Xsubi}{\ensuremath{X_{l}}}

\newcommand{\Nevent}{\ensuremath{N_\text{event}}}

\begin{document}

\title{Cost-efficient Bayesian emulation of quark-gluon plasma modeling with variable statistical precision}

\author{R.~Ehlers}
\affiliation{Nuclear Science Division, Lawrence Berkeley National Laboratory, Berkeley CA 94270.}
\affiliation{Department of Physics, University of California, Berkeley CA 94270.}

\author{Y.~Ji}
\affiliation{Department of Statistical Science, Duke University, Durham NC 27708.}
\affiliation{JMP Statistical Discovery LLC, Cary NC 27513.}

\author{P.~M.~Jacobs}
\affiliation{Nuclear Science Division, Lawrence Berkeley National Laboratory, Berkeley CA 94270.}
\affiliation{Department of Physics, University of California, Berkeley CA 94270.}

\author{S.~Mak}
\affiliation{Department of Statistical Science, Duke University, Durham NC 27708.}

\date{\today}

\begin{abstract}

We present VarP-GP, a new cost-efficient Bayesian emulator for expensive computational models with variable statistical precision. We focus on the interpretation of measurements of the quark-gluon plasma (QGP) generated in high-energy nuclear collisions, through comparison to numerical models using Bayesian Inference. Such inference calculations are computationally expensive and require surrogate model emulation, which is commonly implemented using Machine Learning (ML)--based Gaussian processes (GPs). Emulator training data are generated by Monte Carlo simulations whose numerical precision depends on the computational resources utilized; improved precision entails greater computational cost. This study utilizes JETSCAPE simulations of inclusive hadron and jet measurements in nuclear collisions at RHIC and the LHC. The VarP-GP emulator combines information from multiple simulation runs with varying precision across the model parameter space, taking advantage of the smoothness in that space of QCD-driven processes.  Comparison to a traditional emulator approach shows a marked reduction in emulator uncertainty at fixed computational cost, indicating that knowledge of the overall contours of the parameter design space is more important for precise emulation than detailed information at a more limited number of design points. As an initial application of VarP-GP, a computationally-expensive model parameter sensitivity study of jet quenching data is reported. The VarP-GP emulator enables new multi-model and many-observable calibrations of QGP data and modeling, which would otherwise not be possible with achievable computing resources.

\end{abstract}

\maketitle

\section{Introduction}\label{sec:introduction}

Strongly--interacting matter at very high temperature constitutes a Quark--Gluon Plasma (QGP), in which quarks and gluons are deconfined and  propagate over distances much large than hadronic dimensions~\cite{Busza:2018rrf,Harris:2024aov}. A QGP filled the universe a few micro--seconds after the Big Bang, and QGP is created and studied today in collisions of energetic heavy atomic nuclei at the Relativistic Heavy Ion Collider (RHIC) and the Large Hadron Collider (LHC)~\cite{STAR:2005gfr,PHENIX:2004vcz,ALICE:2022wpn,CMS:2024krd}. 

Experimental measurements at RHIC and the LHC are not directly interpretable in terms of the structure and dynamics of the QGP; rather, QGP properties must be inferred from the comparison of collider data with theoretical calculations and their implementation in detailed numerical models of high--energy nuclear collisions~\cite{Busza:2018rrf,Harris:2024aov}. Such comparisons have revealed that the QGP exhibits collective behavior, flowing as a near-perfect liquid with the smallest specific shear viscosity allowed by Nature~\cite{Busza:2018rrf,Harris:2024aov}, and that the QGP is opaque to energetic quarks and gluons (``jets''). Measurements of the in--medium modification of jet properties probe the microscopic structure of the QGP (``jet quenching'')~\cite{Cunqueiro:2021wls,Apolinario:2022vzg,Wang:2025lct}.

Bayes' Theorem \cite{Bayes:1764vd,gelman1995bayesian} provides a rigorous approach to the comparison of complex data and theoretical models, to test their consistency and to infer model parameters. Application of Bayes' Theorem requires the specification of a prior distribution in parameter space, which encodes pre-existing knowledge. Inference utilizes the likelihood, the probability that the data are described by a given choice of model parameters. A systematic scan of parameter space is then performed to determine that subset of the prior distribution for which the physics simulation best describes the data (posterior distribution). 

Inference calculations are computationally demanding if the model calculation in the likelihood function (the ``forward model'') is computationally expensive, or if the parameter space dimensionality is large. Artificial Intelligence and Machine Learning (AI/ML) tools can improve the computational efficiency of the inference process, in some cases rendering extremely expensive calculations tractable with achievable High--Performance Computing (HPC) resources.

\begin{figure*}[hbt!]
    \centering
    \includegraphics[width=0.8\linewidth]{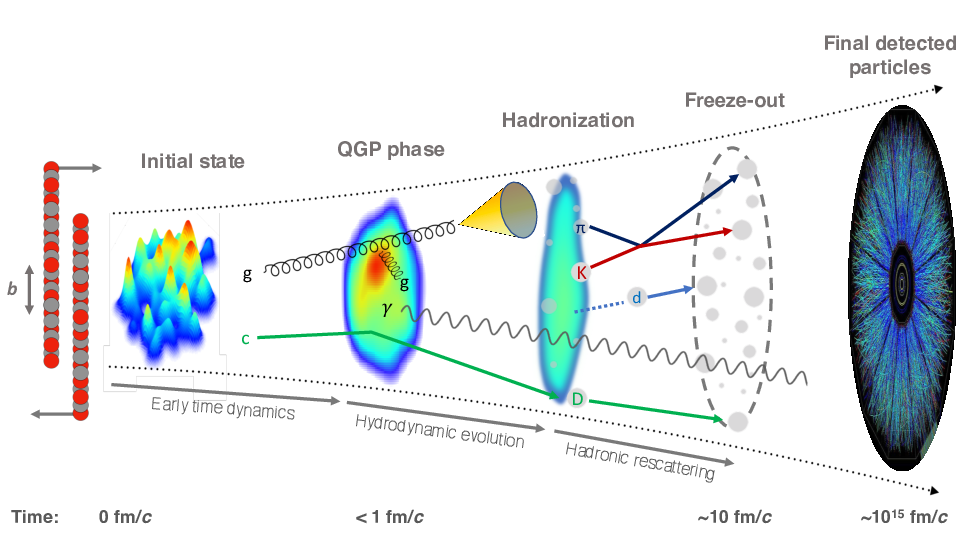}
    \caption{Schematic illustration of the stages of evolution of a high--energy heavy--ion collision which generates a Quark--Gluon Plasma. Figure from~\cite{Arslandok:2023utm}.}
    \label{fig:HIC}
\end{figure*}

Figure~\ref{fig:HIC} shows a prototypical computationally-expensive forward model calculation~\cite{Arslandok:2023utm}: the generation, evolution, and decay of QGP generated in a heavy--ion collision. This multi--step Monte Carlo model incorporates sequential calculations of parton distributions of the colliding nuclei (``initial state''); hard (high momentum-transfer \qsqr) process such as jets, heavy--flavor quarks, and direct photons; QGP equilibration and hydrodynamic evolution; hadron formation; and chemical and kinetic freeze--out. The viscous relativistic hydrodynamic evolution of the QGP is computationally the most demanding step in this sequence. A recent calculation using a two-dimensional approximation required 40 CPU--minutes at an HPC facility to simulate a single central \PbPb\ collision at LHC energies, while the full three-dimensional calculation required 15 CPU--hours~\cite{CShenPrivate}. 

Models of QGP dynamics typically have between 5 and 20 parameters, with inference requiring 200--400 ``design points'' in the prior parameter space. An ML-based emulator is trained on the calculations at the design points, and is then used to predict values and uncertainties of experimental observables continuously over the parameter space. Bayesian Inference employs the trained emulator with Markov Chain Monte Carlo sampling to explore the parameter space~\cite{gelman1995bayesian}. In particular, Gaussian process (GP) emulators are widely used in scientific and engineering applications~\cite{miller2024expected,chen2021function}, and have been applied successfully in Bayesian Inference studies of the QGP~\cite{JETSCAPE:2020mzn, JETSCAPE:2020shq, ji2023graphical,ji2024conglomerate,li2025additive}.

As a recent example, the first multi-observable Bayesian Inference analysis of jet quenching data, which was carried out on an HPC facility, required $\mathcal{O}$(10k) core-hours for each design point in a 6-dimensional prior parameter space~\cite{JETSCAPE:2024cqe}. The calculation employed an ML-based Active Learning algorithm~\cite{cohn1996active} for dynamic selection of parameter--space points, to improve computational efficiency. The full calculation sampled $\approx200$ design points in the prior parameter space, requiring a total of 5M CPU-core hours~\cite{JETSCAPE:2024cqe}. Even with the efficiency gain provided by ML tools, the computational effort for such calculations approaches the limit of achievable HPC resources.

Emulators in such calculations require good precision at each design point; specifically, at each design point the statistical error of the simulation should be less than the corresponding systematic uncertainty and statistical error of the experimental data being compared. This requirement has typically been met by ensuring that the same high relative statistical precision is achieved at every design point, corresponding to the highest-precision experimental data being compared. Such \textit{uniform} statistical precision at all design points is termed ``homoskedastic'' in the statistics literature \cite{gramacy2020surrogates}. However, for Bayesian Inference calculations incorporating heterogeneous datasets with widely varying experimental precision, the homoskedastic requirement may be unnecessarily conservative, resulting in inefficient utilization of available computing resources for emulator training. 

Such inefficiencies indeed arise in QGP inference computations, because different parameter sets emphasize different aspects of QGP physics and are thereby sensitive to different experimental data, which have widely varying relative precision. The most commonly used observables to constrain jet quenching model parameters are the inclusive production of hadrons and jets~\cite{Wang:2025lct,JETSCAPE:2024cqe}, whose data from RHIC and the LHC have widely different experimental precision at low and high transverse momentum (\pT) for hadron yield measurements, and overall markedly better experimental precision for hadrons than for reconstructed jets~\cite{Cunqueiro:2021wls,Apolinario:2022vzg}. 

Optimal utilization of HPC resources for these computations requires dynamical tuning of the target statistical precision of the simulations, based upon the specific characteristics of each design point. Furthermore, the smooth evolution of simulated distributions as a function of the model parameters can be utilized to reduce precision requirements for design points that are nearby in parameter space. The property of
\textit{variable} target statistical precision at different design points is termed ``heteroskedastic'' in the statistics literature \cite{gramacy2020surrogates}. 

This paper presents the Variable Precision GP (VarP-GP) model, a new heteroskedastic GP emulator that is trained on calculations that have variable statistical precision across the parameter design space. The VarP-GP emulator leverages two interacting GP models that jointly learn the mean observable response and varying precision behavior over the parameter space. This permits the ``pooling'' of information from high--precision calculations, enabling accurate emulation at reduced computational efforts. This heteroskedastic emulation performance is enhanced by a novel experimental design procedure that selects the target event counts, and thereby target statistical precision, at each design point. The performance of VarP-GP is assessed by comparing it to that of a conventional GP emulator, using a representative selection of experimental measurements.

Our heteroskedastic VarP-GP emulator has two notable advantages in comparison to conventional homoskedastic GP emulation:

\begin{itemize}
\item Pooling of information from high-precision calculations over the parameter space, which reduces the precision required at design points that are near high-precision points. This provides computational savings to achieve comparable emulator accuracy, or equivalently, improves emulation performance for a fixed computational cost budget. 
\item No requirement of similar precision at all design points, which is complex to implement given the broad variation in simulation performance over the design space.
\end{itemize}

The paper is structured as follows: Sect.~\ref{sect:hetgp} provides background context and presents the VarP-GP emulation framework; Sect.~\ref{sect:physicsModelBayesianAnalysis} describes the QGP physics model and experimental observables used in the study; Sect.~\ref{sect:EmulatorSetupData} outlines the emulator setup and data employed;
Sect.~\ref{sect:results} evaluates the emulation accuracy and computational costs of the VarP-GP, and reports sensitivity studies; and Sect.~\ref{sect:summary} presents conclusions.

\section{Emulation with Variable Statistical Precision}
\label{sect:hetgp}

We first review the conventional GP emulator, which is trained using design points with constant statistical precision. We then present the statistical modeling framework and training algorithm for our VarP-GP emulator, and outline an experimental design approach for selecting the target event counts.

\begin{figure}[hbt!]
    \centering
    \includegraphics[width=\linewidth]{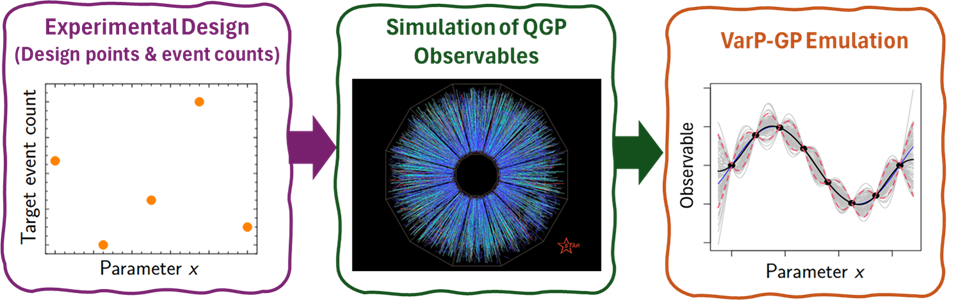}
    \caption{Schematic visualization of the VarP-GP workflow.}
    \label{fig:hetGPWorkflow}
\end{figure}

\subsection{Overview of GP Emulation}

Consider a calculation that is carried out at a set of design points $\mathbf{x}_1, \cdots, \mathbf{x}_n$ over a parameter space $\mathcal{X} \subseteq \mathbb{R}^d$, where $d$ is the number of parameters in the calculation. For design point $\mathbf{x}_i$, let $y_i$ denote its simulated output. Existing emulators in the QGP literature largely assume such outputs follow the generative model:
\begin{equation}
y_i = f(\mathbf{x}_i) + \epsilon_i, \quad \epsilon_i \overset{i.i.d.}{\sim} \mathcal{N}(0,1/p).
\label{eq:const}
\end{equation}
Here, $f(\mathbf{x}_i)$ denotes the mean observable at design point $\mathbf{x}_i$, $\epsilon_i$ denotes its statistical noise, and $\mathcal{N}(0,s^2)$ denotes a zero-mean Gaussian distribution with variance $s^2$. In Eq.~\eqref{eq:const}, the statistical noise at each design point is homoskedastic; it follows an independent and identically distributed ($i.i.d.$) Gaussian distribution with \textit{constant} statistical precision $p$.

An emulator model uses these simulated observables as training data to provide a probabilistic prediction of $f(\mathbf{x}_{\rm new})$ at a new parameter point $\mathbf{x}_{\rm new}$. A GP is typically employed as the prior distribution on the unknown response surface $f$, denoted $f(\cdot) \sim \text{GP}\{\mu,k(\cdot,\cdot)\}$, where $\mu$ is its mean parameter and $k(\cdot,\cdot)$ is its covariance kernel. Common choices of the covariance kernel are the squared-exponential and Mat\'ern kernels \cite{gramacy2020surrogates}. The GP generates a prior stochastic process in the space of plausible functions for $f$, with function smoothness controlled by the kernel $k$.

Emulator predictions are then made by conditioning on the simulated observables $\mathbf{y} = (y_1, \cdots, y_n)$. Suppose we wish to predict $f(\mathbf{x}_{\rm new})$ at a new parameter point $\mathbf{x}_{\rm new}$. For the homoskedastic model \eqref{eq:const}, the posterior distribution of $f(\mathbf{x}_{\rm new})$ conditional on $\mathbf{y}$ takes the Gaussian form~\cite{gramacy2020surrogates}:
\begin{equation}
[f(\mathbf{x}_{\rm new})|\mathbf{y}] \sim \mathcal{N}\{\mu(\mathbf{x}_{\rm new}),\sigma^2(\mathbf{x}_{\rm new})\},
\label{eq:gpconst1}
\end{equation}
where:
\begin{align}
\begin{split}
\mu(\mathbf{x}) &= \mu + \mathbf{k}(\mathbf{x})^\top (\mathbf{K} + p^{-1}\mathbf{I}_{n \times n})^{-1} (\mathbf{y} - \mu \mathbf{1}_n),\\
\sigma^2(\mathbf{x}) &= k(\mathbf{x},\mathbf{x})-\mathbf{k}(\mathbf{x})^\top (\mathbf{K} + p^{-1}\mathbf{I}_{n \times n})^{-1}\mathbf{k}(\mathbf{x}).
\label{eq:gpconst2}
\end{split}
\end{align}
Here, $\mathbf{k}(\mathbf{x}) = [k(\mathbf{x}_i,\mathbf{x})]_{i=1}^n$, $\mathbf{K} = [k(\mathbf{x}_i, \mathbf{x}_j)]_{i,j=1}^n$, $\mathbf{I}_{n \times n}$ is an $n \times n$ identity matrix, and $\mathbf{1}_n$ is a length-$n$ vector of ones. Equation~\eqref{eq:gpconst1} provides the basis for probabilistic emulation: its posterior mean $\mu(\cdot)$ serves as a predictor of $f$ over the parameter space, and its posterior variance $\sigma^2(\cdot)$ quantifies its emulation uncertainty. The closed-form nature of the above posterior distribution is key for efficient Bayesian Inference of the QGP via Markov Chain Monte Carlo techniques (see Ref~\cite{JETSCAPE:2020mzn}). Model parameters in the GP can be learned via maximum likelihood estimation or in a fully Bayesian manner \cite{gramacy2020surrogates}.

\subsection{VarP-GP Emulator}

Let $y_1, \cdots, y_n$ again denote the simulated observables at design points $\mathbf{x}_1, \cdots, \mathbf{x}_n$, and let $m_i$ denote the simulated event count for $y_i$. Our VarP-GP adopts the following generative model on the simulated outputs $y_1, \cdots, y_n$:
\begin{equation}
y_i = f(\mathbf{x}_i) + \epsilon_i, \quad \epsilon_i \overset{i.i.d.}{\sim} \mathcal{N}\left\{0,\frac{s^2(\mathbf{x}_i)}{m_i}\right\}.
\label{eq:het}
\end{equation}
\noindent
In the above, the precision of the simulated observable $y_i$ is modeled as $p_i = m_i / s^2(\mathbf{x}_i)$, which has two important properties. First, as the simulated event count $m_i$ increases, the statistical precision of $y_i$ should increase linearly with $m_i$; equivalently, its statistical variance should decrease proportionally to $m_i$. 
Here, the $s^2(\mathbf{x}_i)$ term models the ``baseline'' statistical variance of the simulated observable for a single event, with $1/s^2(\mathbf{x}_i)$ specifying its ``baseline'' statistical precision.
This precision corresponds to a measured physical yield, such as a cross section.
Second, unlike the homoskedastic model (Eq.~\eqref{eq:const}) used widely in QGP emulation, the elements of the precision vector $(p_1, \cdots, p_n)$ can vary due to a differing number of events $m_i$ or from a differing baseline statistical precision $1/s^2(\mathbf{x}_i)$ at each design point\footnote{Baseline statistical precision at each design point can vary due to the underlying physical model. For example, a stronger coupling constant will lead to increased partonic showering, which results in increased yield at fixed momentum.}.

Since both the observable mean $f(\cdot)$ and baseline statistical variance $s^2(\cdot)$ functions are unknown, we assign them independent GP priors of the following form:
\begin{equation}
f(\cdot) \sim \text{GP}\{\mu_f,k_f(\cdot,\cdot)\}, \; \log s^2(\cdot) \sim \text{GP}\{\mu_s,k_s(\cdot,\cdot)\}.
\label{eq:hetgp}
\end{equation}
The joint GPs on $f(\cdot)$ and $\log s^2(\cdot)$ provide a statistical framework for pooling information from high-fidelity calculations over the parameter space $\mathcal{X}$, by learning how the baseline statistical precision varies over $\mathcal{X}$. Here, the log-transform on $s^2(\cdot)$ ensures that the modeled statistical precision is positive. The emulator model in Eq.~\eqref{eq:hetgp} is related in form to the heteroskedastic statistical model in Ref.~\cite{binois2021hetgp}, although the latter was not used in the context of emulation.

For the GP kernel $k_f$, we adopt the widely-used Mat\'ern covariance kernel \cite{stein1999interpolation} of the form:
\small
\begin{align}
k_f(\mathbf{x},\mathbf{x}') &= \sigma^2_f \left( 
1 + \sqrt{5} \sqrt{ \sum_{l=1}^d \frac{(x_l - x_l')^2}{\theta_{f,l}^2} } 
+ \frac{5}{3} \sum_{l=1}^d \frac{(x_l - x_l')^2}{\theta_{f,l}^2} 
\right) \\
& \quad \quad \times \exp\left( -\sqrt{5} \sqrt{ \sum_{l=1}^d \frac{(x_l - x_l')^2}{\theta_{f,l}^2} } \right), \notag
\end{align}
\normalsize
where $\sigma_f^2$ and $\boldsymbol{\theta}_f = \{\theta_{f,1}, \cdots, \theta_{f,d}\}$ are its scale and 
length-scale parameters. The scale parameter $\sigma_f^2$ controls the function scale of $f$, and the length-scale parameters $\boldsymbol{\theta}_f$ control the influence of each of the $d$ parameters in the parameter space. A similar Mat\'ern kernel is used for $k_s$, with scale and length-scale parameters $\sigma_s^2$ and $\boldsymbol{\theta}_s = \{\theta_{s,1}, \cdots, \theta_{s,d}\}$.

Two additional model developments are needed for effective emulation of QGP observables. First, experimental collider data are reported as ensemble-averaged distributions, i.e., distributions whose mean and statistical precision are determined by averaging over a large population of collider events. In contrast, current heteroskedastic models \cite{binois2018practical} require the training data in replicated form, i.e., the set of events themselves. To bridge this gap, a different training algorithm is therefore needed, which we present in the following. Second, the choice of simulated event counts $m_1, \cdots, m_n$ should be carefully optimized for effective pooling of information from high-fidelity calculations; we will address this later in Sect.~\ref{sec:design}.

Model training of VarP-GP requires tuning of the model hyperparameters $\boldsymbol{\Theta} = \{\mu_f,\mu_s,\sigma^2_f,\sigma^2_s,\boldsymbol{\theta}_f,\boldsymbol{\theta}_s\}$ from data. An effective approach for such tuning is maximum likelihood estimation \cite{gramacy2020surrogates}, which selects $\boldsymbol{\Theta}$ as the hyperparameters which maximize the likelihood of the training data. Such data include the simulated observables $\mathbf{y} = (y_1, \cdots, y_n)$ and their corresponding precisions $\mathbf{p} =(p_1, \cdots, p_n)$.
As shown in Appendix~\ref{app:deriv}, the log-likelihood function of such data takes the form:
\begin{align}
\begin{split}
l(\boldsymbol{\Theta}) &= -\frac{1}{2}\log|\mathbf{K}_f+\mathbf{P}^{-1}| - \frac{1}{2}\log|\mathbf{K}_s| \\
& \quad \quad -\frac{1}{2}(\mathbf{y}-\mu_f\mathbf{1}_n)^T(\mathbf{K}_f+\mathbf{P}^{-1})^{-1}(\mathbf{y}-\mu_f\mathbf{1}_n) \\
&\quad \quad - \frac{1}{2}(\log \mathbf{s}^2-\mu_s\mathbf{1}_n)^T\mathbf{K}_s^{-1}(\log \mathbf{s}^2-\mu_s\mathbf{1}_n) + C.
\end{split}
\label{eq:lkhd}
\end{align}
Here, $\mathbf{K}_f = [k_f(\mathbf{x}_i,\mathbf{x}_j)]_{i,j=1}^n$ and $\mathbf{K}_s = [k_s(\mathbf{x}_i,\mathbf{x}_j)]_{i,j=1}^n$ are the GP covariance matrices for $f$ and $\log s^2$, respectively, $\mathbf{P} = \text{diag}\{ p_1, \cdots, p_n \}$ is the diagonal matrix of statistical precisions for design points, $\mathbf{s}^2 = (m_1/p_1, \cdots, m_n/p_n)$, and $C$ captures remaining constant terms in $\boldsymbol{\Theta}$.

Note that Eq.~\eqref{eq:lkhd} is a function of the model hyperparameters $\boldsymbol{\Theta}$, which we aim to optimize. This can be carried out using a maximum likelihood approach \cite{casella2024statistical}, by maximizing $l$ with respect to $\boldsymbol{\Theta}$ via numerical optimization algorithms. We recommend the use of the low-memory BFGS quasi-Newton algorithm \cite{nocedal2006numerical} for solving this optimization problem. Since each evaluation of $\ell$ requires $\mathcal{O}(n^3)$ work, this optimization can be efficiently performed provided the number of design points $n$ is not too large (e.g., $n \leq 1,000$). For larger values of $n$, scalable GP techniques (e.g., inducing points \cite{titsias2009variational,li2025prospar}) can be used to accelerate hyperparameter optimization.

VarP-GP emulator predictions are then generated as follows. For the heteroskedastic model \eqref{eq:het}, utilizing Eqs.~\eqref{eq:gpconst1} and \eqref{eq:gpconst2}, it can be shown that the posterior predictive distribution of the mean observable $f(\mathbf{x}_{\rm new})$ at a new parameter point $\mathbf{x}_{\rm new}$ takes the form:
\begin{equation}
[f(\mathbf{x}_{\rm new})|\mathbf{y},\mathbf{p}] \sim \mathcal{N}\{ \mu_f(\mathbf{x}_{\rm new}), \sigma^2_f(\mathbf{x}_{\rm new}) \},
\label{eq:gphet1}
\end{equation}
where:
\small
\begin{align}
\begin{split}
\mu_f(\mathbf{x}) &= \mu_f + \mathbf{k}_f(\mathbf{x})^\top (\mathbf{K}_f + \mathbf{P}^{-1})^{-1} (\mathbf{y} - \mu_f \mathbf{1}_n),\\
\sigma^2_f(\mathbf{x}) &= k_f(\mathbf{x},\mathbf{x})-\mathbf{k}_f(\mathbf{x})^\top (\mathbf{K}_f + \mathbf{P}^{-1})^{-1}\mathbf{k}_f(\mathbf{x}).
\label{eq:gphet2}
\end{split}
\end{align}
\normalsize
Here, $\mathbf{k}_f(\mathbf{x})$ is the analogue of $\mathbf{k}(\mathbf{x})$ in Eq.~\eqref{eq:gpconst2} using kernel $k_f$, and $\mathbf{K}_f$ has been defined previously. Equations~\eqref{eq:gphet1} and \eqref{eq:gphet2} serve as the basis for VarP-GP emulation: $\mu_f(\mathbf{x}_{\rm new})$ gives the predicted observable at a new point $\mathbf{x}_{\rm new}$, and $\sigma_f^2(\mathbf{x}_{\rm new})$ quantifies its emulation uncertainty in the form of its posterior predictive variance.

One can further predict the baseline statistical log-variance (or equivalently, the baseline statistical precision via an appropriate transformation) at a new parameter point $\mathbf{x}_{\rm new}$.
Such a prediction provides insight into the behavior of the physical model, establishing a link between model parameters and observable yields and uncertainties, which may enable additional precision target optimization\footnote{For example, we could predict the change in precision for a fixed number of generated events as the value for switching between high and low virtuality models varies. Improved precision for some design points would enable physically motivated reductions of target event counts, lowering computational cost.}.
Under the VarP-GP model, the posterior distribution of $\log s^2(\mathbf{x}_{\rm new})$ can be shown to take the form $[\log s^2(\mathbf{x}_{\rm new}) |\mathbf{y},\mathbf{p}] \sim \mathcal{N}\{{\mu}_s(\mathbf{x}_{\rm new}), {\sigma}^2_s(\mathbf{x}_{\rm new})\}$, where:
\begin{align}
\begin{split}
{\mu}_s(\mathbf{x}) &= \mu_s + \mathbf{k}_s(\mathbf{x})^\top \mathbf{K}_s^{-1}(\log \mathbf{s}^2 - \mu_s \mathbf{1}_n),\\
{\sigma}^2_s(\mathbf{x}) &= k_s(\mathbf{x}, \mathbf{x}) - \mathbf{k}_s(\mathbf{x})^\top \mathbf{K}_s^{-1} \mathbf{k}_s(\mathbf{x}).
\end{split}
\label{eq:gphet3}
\end{align}
Here, ${\mu}_s(\mathbf{x}_{\rm new})$ gives the predicted baseline log-variance at the new point, and ${\sigma}^2_s(\mathbf{x}_{\rm new})$ quantifies its uncertainty. Appendix \ref{app:deriv} provides the derivation on these posterior distributions.

\subsection{Experimental Design}
\label{sec:design}

To enhance the cost--effectiveness of heteroskedastic emulation, VarP-GP should be coupled with a strategic choice of design points and their corresponding target event counts that control statistical precision. We address this in three steps: (i) the selection of the design points $\mathbf{x}_1, \cdots, \mathbf{x}_n$, (ii) the selection of a candidate set for target event counts $\mathcal{M}$, and (iii) the ``pairing'' of selected design points to event counts in $\mathcal{M}$. 

Step (i) can be performed with existing techniques. We recommend setting the number of design points $n$ between $5d$ and $10d$, following the rule-of-thumb proposed for GP emulation~\cite{loeppky2009choosing,mak2018efficient}. The design points $\mathbf{x}_1, \cdots, \mathbf{x}_n$ can then be selected from a Latin hypercube design (LHD; \cite{mckay2000comparison}), which ensures that such points are spread over the parameter space, particularly on one-dimensional projections. LHDs have been successfully employed in recent QGP analyses; see, e.g., \cite{JETSCAPE:2024cqe,JETSCAPE:2020mzn,JETSCAPE:2020shq}.

Next, Step (ii) involves the selection of a candidate set $\mathcal{M}$, from which the target event counts $m_1, \cdots, m_n$ are chosen. Let $\underline{m}$ and $\overline{m}$ be the pre-specified lower and upper bounds of the target event count at a given design point. Such bounds can be specified based on physics expectations. The upper bound $\overline{m}$ can be set to the maximum desired event target, as determined by comparing to the maximum precision of measured observables. The lower bound $\underline{m}$ is set through a balance between competing constraints: the target must be sufficiently high to carry some minimum physical information, but should be set as low as possible to maximize the computational efficiency. For this study, $\underline{m}$ must be at least 30\%  of the maximum event count $\overline{m}$. Note that the event count $m$ can be viewed as a ``fidelity'' parameter, in that a larger $m$ results in a more accurate calculation, and vice versa.

In designing fidelity levels for GP emulation, Ref.~\cite{shaowu2023design} recommends the following candidate set $\mathcal{M}$:
\begin{equation}
\mathcal{M} = \left\{ m : m = \underline{m} \left(\frac{\overline{m}}{\underline{m}}\right)^{u}, u = \frac{i-1}{n-1}, i = 1, \cdots, n \right\}.
\label{eq:cand}
\end{equation}
This choice of $\mathcal{M}$ has advantageous properties for heteroskedastic emulation. First, it ensures an even spread of simulated precision levels, where some design points have low precision and others have high precision. Second, event counts in this candidate set are inversely proportional in $m$, such that there are more candidate precision levels near the lower bound $\underline{m}$ than the upper bound $\overline{m}$. This enhances emulator cost efficiency, since it avoids running many design points at high precision to reduce computational burden.

Finally, given the selected design points and the event count candidate set, Step (iii) pairs each design point with its corresponding event count from the candidate set $\mathcal{M}$. Recall that a key benefit of VarP-GP is that it pools information from high-precision calculations over the parameter space, thereby reducing the need to run all design points at high precision. To enhance this benefit, the pairing of design points to event counts should be performed so that high-precision runs are spread over the parameter space, i.e., no two design points that are run at high precision are adjacent.

To achieve this for given design points $\mathbf{x}_1, \cdots, \mathbf{x}_n$, our pairing approach selects the event counts $m_1, \cdots, m_n$ to maximize the following criterion:
\begin{equation}
\underset{m_1, \cdots, m_n \in \mathcal{M}}{\text{argmax}} \; \min_{\substack{i,j=1,\cdots,n\\ i \neq j}} \; \|\mathbf{x}_i - \mathbf{x}_j\|_2 \cdot |m_i-m_j|.
\label{eq:pair}
\end{equation}
The intuition is as follows. Suppose that two design points $\mathbf{x}_i$ and $\mathbf{x}_j$ are nearby in the parameter space, i.e., $\|\mathbf{x}_i - \mathbf{x}_j\|_2$ is small. For such nearby points, we wish to avoid running both at \textit{high} event counts (i.e., large $m_i$ and $m_j$), to ensure that computational resources are not wasted when pooling information from high-precision calculations over the parameter space. We further wish to avoid running both at \textit{low} event counts (i.e., small $m_i$ and $m_j$), to ensure that there are nearby high-precision points to improve emulation accuracy. The pairing criterion in Eq.~\eqref{eq:pair} discourages such scenarios. For nearby design points $\mathbf{x}_i$ and $\mathbf{x}_j$ (i.e., with $\|\mathbf{x}_i - \mathbf{x}_j\|_2$ small), both of the above scenarios result in a small $|m_i-m_j|$, which then yields a small criterion for Eq.~\eqref{eq:pair}. By maximizing the pairing criterion in Eq.~\eqref{eq:pair}, we thereby discourage such scenarios and ensure a varied selection of event counts over the parameter space.

\begin{figure}[hbt!]
    \centering
    \includegraphics[width=0.49\linewidth]{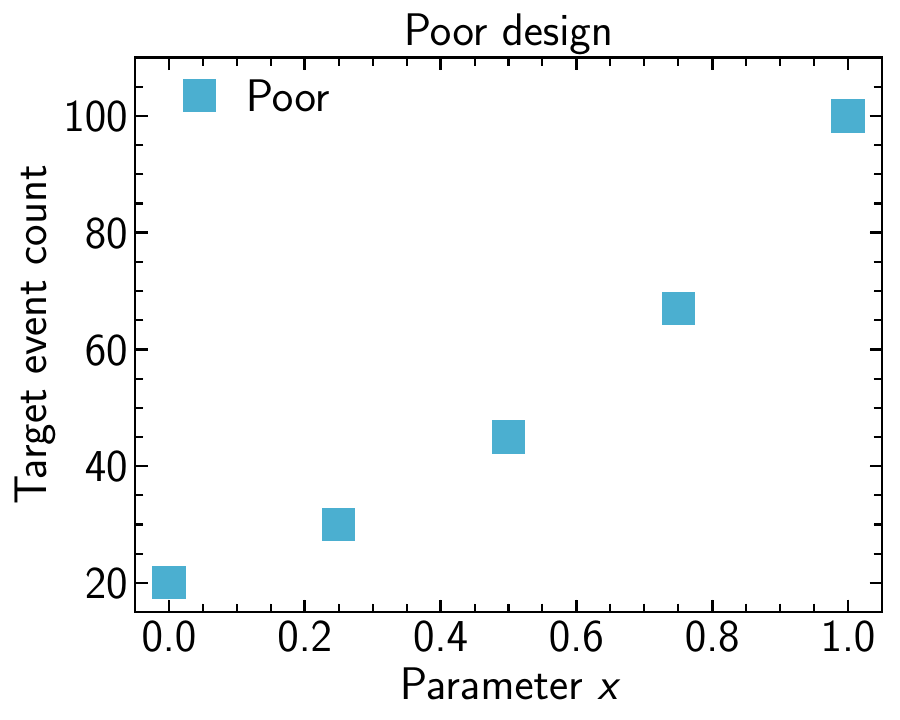}
    \includegraphics[width=0.49\linewidth]{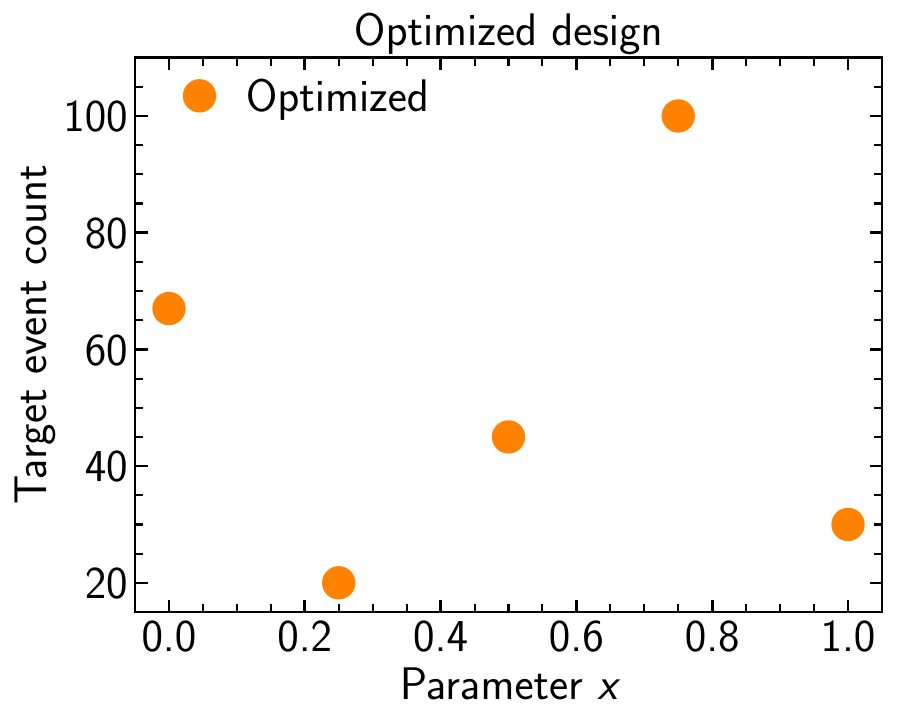}
    \caption{A toy example with $n=5$ design points. (Left) A poor pairing of design points with event counts in $\mathcal{M}$. (Right) An optimized pairing of design points with event counts in $\mathcal{M}$, using the proposed criterion from Eq.~\eqref{eq:pair}.}
    \label{fig:designvis}
\end{figure}

Figure~\ref{fig:designvis} visualizes this experimental design approach with a simple example. Suppose there is a single parameter $x$, and calculations are performed at $n=5$ design points. Following Step (i), points would be placed in an evenly-spaced fashion at $x = \{0, 0.25, 0.5, 0.75, 1\}$. For Step (ii), suppose the lower and upper event counts $\underline{m}$ and $\overline{m}$ are set as 20 and 100, in which case the candidate set is chosen as $\mathcal{M} = \{20, 30, 45, 67, 100\}$. Figure~\ref{fig:designvis} (left) shows a poor pairing of design points with event counts in $\mathcal{M}$, where both high precision and low precision points are grouped together. Figure~\ref{fig:designvis} (right) shows our optimized pairing from Step (iii), where event count selection is more varied over the parameter space.

\section{Physics modeling and experimental observables}
\label{sect:physicsModelBayesianAnalysis}

Numerical modeling of heavy--ion collisions is complex and computationally expensive. Such models are formulated in a Monte Carlo approach with multiple distinct stages, each of which requires calibration and validation. The numerical model in this study is the same as that used in Ref.~\cite{JETSCAPE:2024cqe} for a comprehensive JETSCAPE Bayesian inference analysis of jet quenching data. In this section, we briefly outline the model and experimental measurements on which the VarP-GP emulator is trained. 

\subsection{Physics model}
\label{sect:PhysModel}

The calculation of jet interactions in the QGP itself has two stages: the generation and evolution of bulk matter, followed by the propagation in the medium of hard parton showers. Simulations utilize the modular JETSCAPE event generation framework~\cite{Putschke:2019yrg}, with the physics model described in Refs.~\cite{JETSCAPE:2022jer,JETSCAPE:2024cqe}. 

For bulk matter simulation, the initial nucleon and energy distributions are based on the \TRENTO\ model~\cite{Moreland:2014oya}. Initial evolution is free--streaming \cite{Liu:2015nwa} for a period $\tau_R$, followed by viscous fluid evolution of the QGP based on Israel-Stewart transient hydrodynamics as implemented in the VISHNU code~\cite{Song:2007ux,Shen:2014vra}. A QGP phase--space element is hadronized at switching temperature $T_{SW}$ using the Cooper-Frye algorithm~\cite{Cooper:1974mv}, with subsequent hadronic multiple scattering simulated by \URQMD~\cite{Bass:1998ca,Bleicher:1999xi}. In this study, the bulk--matter simulations use the maximum a posteriori (MAP) parameters from Ref.~\cite{Bernhard:2019bmu}, which generate similar QGP properties to those in Ref.~\cite{JETSCAPE:2020mzn,JETSCAPE:2020shq}. 

For parton shower simulations in nuclear collisions, the geometric distribution of hard-scattering vertices follows the nucleon-nucleon collision distribution in \TRENTO. Initial--state radiation and hard scattering processes are calculated using the \PYTHIA\ generator for \pp\ collisions~\cite{Bierlich:2022pfr}. Final--state radiation, including that due to jet--QGP interactions, is calculated by the virtuality--dependent approach described in Ref.~\cite{JETSCAPE:2024cqe}, in which high--virtuality evolution is treated perturbatively, and low--virtuality evolution is treated non--perturbatively. Hadronization of the jet shower is described in Ref.~\cite{JETSCAPE:2022jer}.
The full set of model parameters are defined in Sect.~\ref{sect:SensitivityAnalysis}.

\subsection{Experimental observables}
\label{sect:Observables}

Jets, which are correlated sprays of the hadrons, are the remnants of energetic quarks and gluons generated by hard scatterings in high--energy collisions~\cite{Sterman:1977wj}. Jets play a central role in many aspects of collider physics~\cite{Salam:2010nqg}. Jets are used to probe the QGP, through medium-induced modification to jet production rates, correlations, and internal jet structure (jet quenching)~\cite{Cunqueiro:2021wls,Apolinario:2022vzg,Wang:2025lct}.

Jet quenching is commonly explored using the inclusive production of high--\pT\ hadrons and jets. High--\pT\ hadrons arise from the fragmentation of a jet into a single hadron which carries a large fraction of the total jet energy. They provide an important proxy for jets in jet quenching studies, since they are more easily measurable than reconstructed jets in the complex background of heavy--ion collisions~\cite{Jacobs:2004qv}. However, techniques to measure reconstructed jets in the heavy--ion collision environment have also been developed, and extensive jet quenching measurements based on reconstructed jets are also available at both RHIC and the LHC~\cite{Cunqueiro:2021wls,Apolinario:2022vzg,Wang:2025lct}. 

Jet quenching effects in inclusive production are quantified by the experimentally measurable ratio \RAA~\cite{Jacobs:2004qv},

\begin{equation}
\RAA=\frac{\dNAA/\dpT}{\TAA\cdot\dsigpp/\dpT},
\label{eq:RAA}
\end{equation}

\noindent
where $\dNAA/\dpT$ is the inclusive hadron or jet yield per event measured in nuclear collisions, and $\dsigpp/\dpT$ is the cross section of the same process measured in \pp\ collisions to provide the experimentally determined distribution in the absence of jet quenching. The normalization factor \TAA, which accounts for the geometry of nuclear collisions and calculated using Glauber modeling~\cite{Jacobs:2004qv,Miller:2007ri}, is defined so that $\RAA=1$ if there are no nuclear--induced modifications, i.e. no net modification of the inclusive production rate due to initial--state nuclear shadowing and final--state jet quenching. Extensive hadron and jet measurements at RHIC and the LHC find values of \RAA\ significantly less than unity, far smaller than can be accounted for by initial--state nuclear shadowing. These observations provide direct evidence of medium-induced energy loss, i.e., jet quenching~\cite{Cunqueiro:2021wls,Apolinario:2022vzg,Wang:2025lct}.

\section{Emulator set-up}
\label{sect:EmulatorSetupData}

An emulator provides an interpolation of simulated data (in this study, hadron and jet \RAA) between design points, continuously over the design parameter space. The goal of this study is to assess the performance of the VarP-GP emulator in a representative physics calculation, not to carry out a comprehensive inference calculation. For clarity and flexibility, we therefore utilize only a small subset of available jet and hadron \RAA\ data that were used in Ref.~\cite{JETSCAPE:2024cqe}: inclusive hadron \RAA\ measured by the CMS collaboration~\cite{CMS:2016xef} and inclusive jet \RAA\ with resolution parameter $R=0.4$ measured by the ATLAS collaboration~\cite{ATLAS:2018gwx}, in \PbPb\ and \pp\ collisions at $\sqrtsNN=5.02$ TeV. This choice nevertheless represents the observables and datasets that will be included in a full multi-observable Bayesian Inference analysis, providing complementary sensitivity to the underlying physics and validating that the methods developed here apply to multi-observable analyses. The data are employed in two distinct ways: as guidance to the dynamic range in \pT\ and \RAA\ of a representative calculation to compare different emulators (Sect.~\ref{sect:EmulatorPerformance}); and as reference for an initial physics application of VarP-GP, to explore parameter sensitivities of the posterior distributions in Ref.~\cite{JETSCAPE:2024cqe} (Sect.~\ref{sect:SensitivityAnalysis}).

For emulator training, following Step (i) in Sect.~\ref{sec:design}, we begin by selecting $n=40$ Latin hypercube design points. Here, the number of design points $n$ is within the recommended range $n = 5-10d$, where $d = 6$ is the number of parameters in the QGP model (see Sect.~\ref{sect:SensitivityAnalysis}). Simulations are then carried out using the physics model (Sect.~\ref{sect:PhysModel}), with 1.25M events simulated  at each design point. After simulation, 12 design points are excluded due to large fluctuations, which indicates poor simulation data quality. With these excluded, the training set has a total of $n=28$ design points. This simulated dataset is then used to train the VarP-GP, HF-GP, and LF-GP emulators, as discussed below.

The following emulators are implemented for comparison as a function of the simulated event budget \Nevent{} (more on this in Sect.~\ref{sect:EmulatorPerformance}):

\begin{itemize}

\item {\bf Variable Precision GP (VarP-GP):} The VarP-GP emulator allocates the event budget \Nevent\ variably amongst the design points. Following Step (ii) in Sect.~\ref{sec:design}, the candidate set $\mathcal{M}$ in Eq.~\eqref{eq:cand} is determined with $\overline{m} = 1.25$M events, where the event count lower bound $\underline{m}$ is set such that the total event count in $\mathcal{M}$ equals \Nevent. For this study, $\underline{m}$ is required to be at least 30\% of $\overline{m}$ to ensure that the simulation carries sufficient physical information. Finally, following Step (iii) in Sect.~\ref{sec:design}, each of the $n=28$ design points in the full simulated dataset is paired with the event counts in $\mathcal{M}$. 

\item {\bf High Fidelity GP (HF-GP):} For a given choice of \Nevent, the conventional HF-GP emulator allocates resources by running simulations at each design point at the highest precision considered in this study (i.e., $\overline{m}=1.25$M events), with the number of training design points approximately equal to $\Nevent / \overline{m}$. This allocation scheme is similar to that used for the GP emulator in Ref.~\cite{JETSCAPE:2024cqe}. In Sect.~\ref{sect:results}, HF-GP is implemented by randomly selecting training design points from the full simulated dataset. 

\item {\bf Low Fidelity GP (LF-GP):} LF-GP can be viewed as the converse of HF-GP. For a given choice of \Nevent, all $n=28$ design points are utilized, with the number of events at each design point taken as $\Nevent/n$.


\end{itemize} 
\noindent
Note that, for a given choice of \Nevent, the number of design points used for VarP-GP training is larger than that for HF-GP and equal to that for LF-GP. 

The hadron and jet \RAA\ data employed here cover a broad range in \pT~\cite{CMS:2016xef,ATLAS:2018gwx}.
The hadron \RAA\ data consist of 21 data points with $4.8<\pT<400$~GeV/$c$, while the jet \RAA\ data consist of 15 data points with $100<\pT<1000$~GeV/$c$. These \pT\ ranges are narrower than those reported in the experimental measurements to ensure that the simulations lie within the region of applicability of the physics model~\cite{JETSCAPE:2024cqe}.

Since the focus of this study is the characterization of emulator performance, the value of \RAA\ is calculated at each \pT\ value using a separate emulator. Calculations at different \pT\ values are therefore statistically independent, enabling the statistical assessment of emulator performance. 

A different procedure is used in practice for large-scale Bayesian Inference calculations of jet quenching. Physically, the distribution of \RAA\ is a smoothly varying function, with typically logarithmic dependence on \pT~\cite{Cunqueiro:2021wls,Apolinario:2022vzg,Wang:2025lct}. A Principle Component Analysis (PCA) is therefore utilized to reduce the number of distinct emulators required~\cite{Paquet:2023rfd,JETSCAPE:2024cqe}, taking advantage of the smooth variation of \RAA\ vs. \pT\ to reduce the computing resources required and to preserve correlations across \pT.

\section{Results}
\label{sect:results}

\subsection{Emulator performance}
\label{sect:EmulatorPerformance}

Emulator performance is assessed as a function of the number of simulated events \Nevent, which represents the computational budget for training. This budget metric is employed, rather than core-hours, because it shows directly the performance impact of varying statistical precision. Here, emulation performance is quantified by comparing the calculated value of \RAA\ from the VarP-GP, HF-GP or LF-GP emulator to that from a reference high-fidelity computation at 23 test design points\footnote{The validation set was simulated at 30 design points drawn from a Latin Hypercube Design, with 7 points removed due to poor simulation quality.}. The reference simulation at each test design point is run to the highest precision considered in the analysis (i.e., $\overline{m}=1.25$M events). 

A separate VarP-GP, HF-GP, or LF-GP emulator is then trained for each value of \pT. For each value of \pT, the Mean Square Error (MSE) is determined by comparing the value of \RAA\ computed by the emulator to that of the reference calculation at the 23 test design points. Smaller MSE corresponds to higher emulation accuracy, and a narrower distribution in MSE indicates more consistent emulation performance over a range of \pT\ values.

\begin{figure*}[hbt!]
    \centering
    \includegraphics[width=0.495\linewidth]{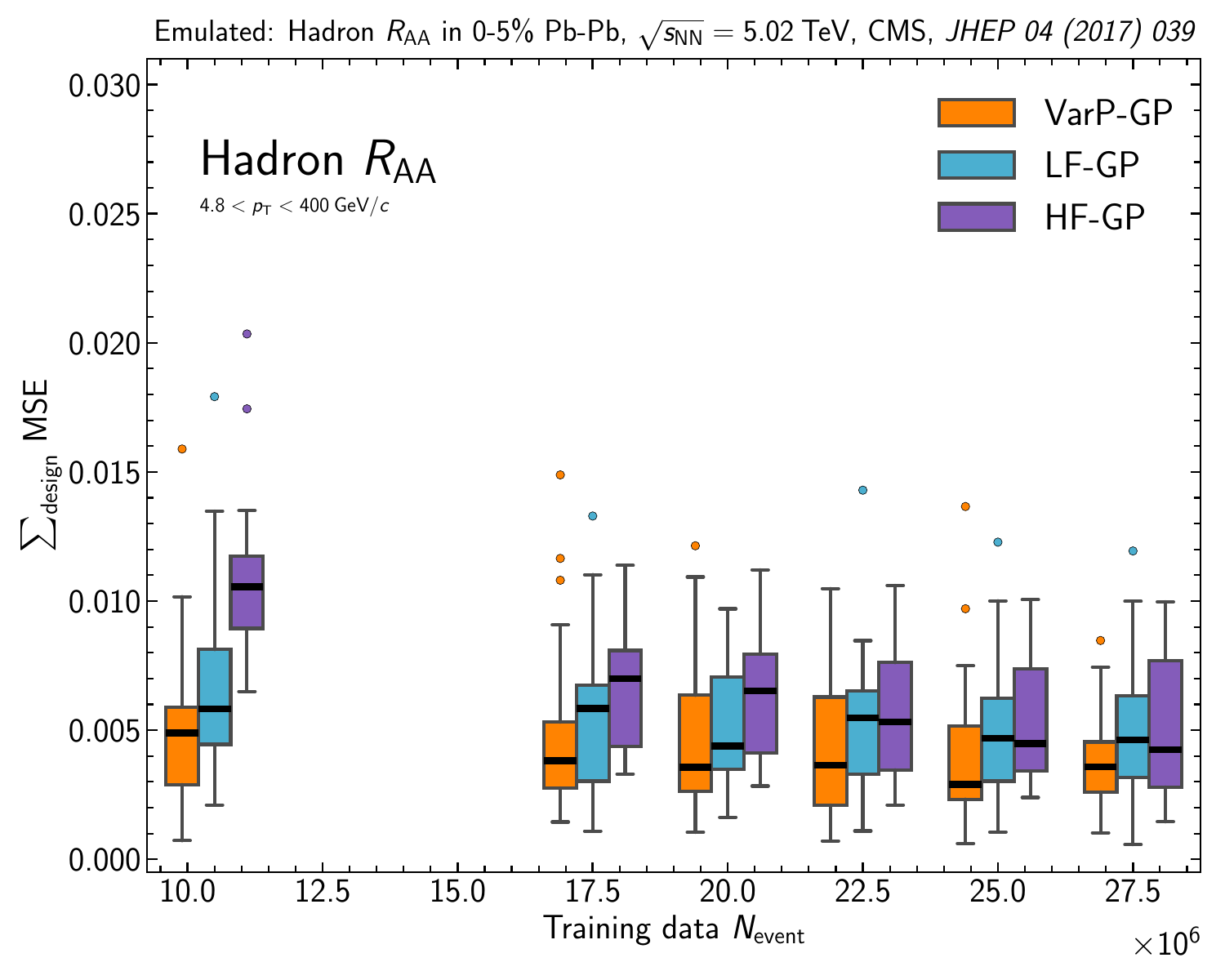}
    \includegraphics[width=0.495\linewidth]{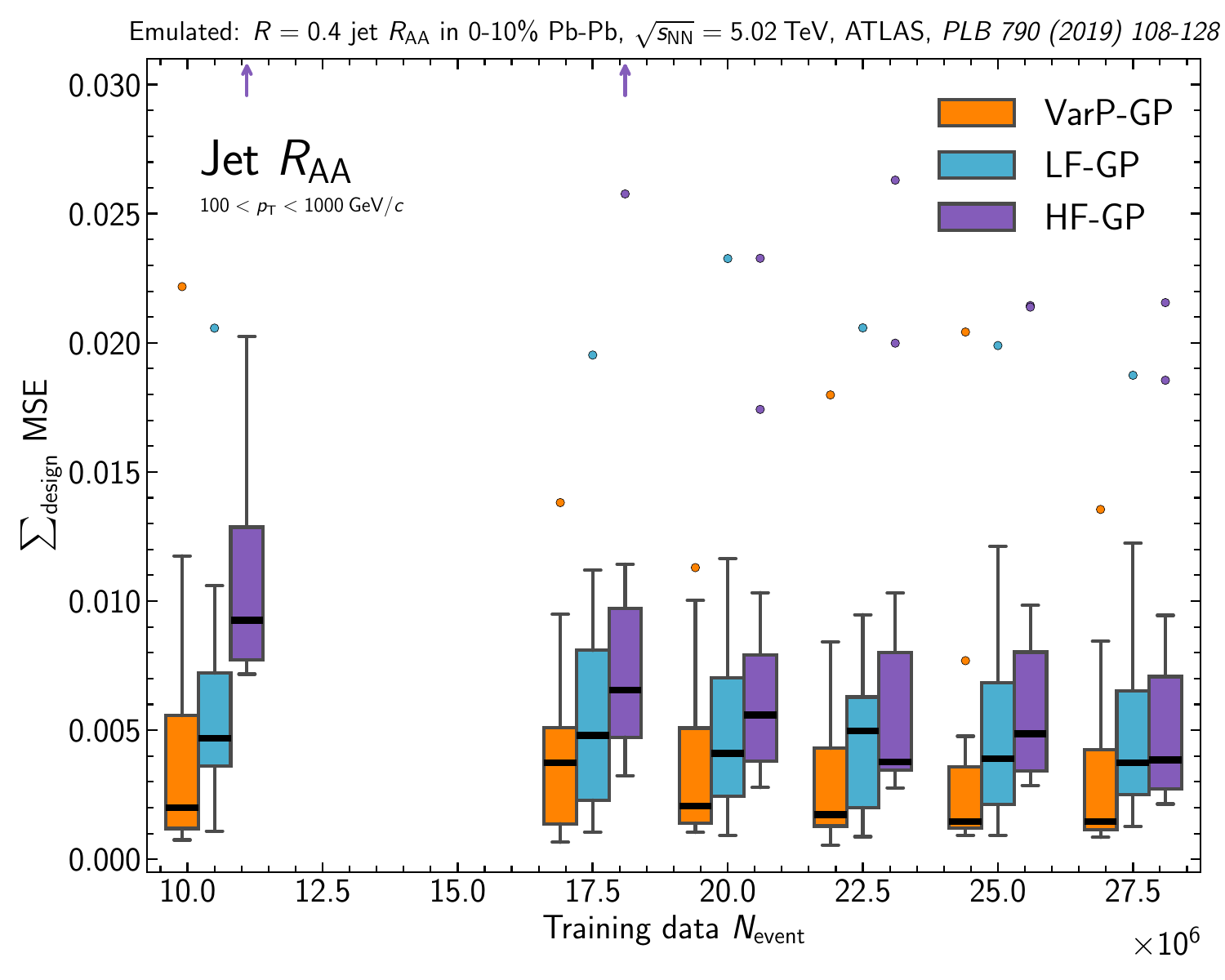}
    \caption{Assessment of VarP-GP, LF-GP, and HF-GP performance as a function of the number of simulated events \Nevent\ for emulator training. For each value of \pT, an MSE is determined by comparing \RAA\ computed by VarP-GP, LF-GP, and HF-GP to that of a reference high-fidelity calculation at 23 test design points. Each figure element shows the distribution of MSE across \pT\ values: thick black line is the median; colored box lower and upper limits are 25\% and 75\% quartiles; and vertical lines beyond the box limits correspond to the largest values within 1.5 times the inter-quartile range. Outliers are shown as individual points. The x-axis values are offset for clarity. Vertical colored arrows at the top of the figure indicate a single MSE value above the maximum y-axis range for a given number of training events. Left: hadron \RAA; right: jet \RAA. Orange: VarP-GP; blue: LF-GP; purple: HF-GP.
}
\label{fig:differentialEmulationPerformance}
\end{figure*}

Figure~\ref{fig:differentialEmulationPerformance} shows the distribution of MSE for the VarP-GP, HF-GP, and LF-GP emulators as a function of \Nevent. The VarP-GP MSE distributions have markedly lower median values and are generally narrower than those of HF-GP over the full range in \Nevent, for both hadron and jet \RAA. Both emulators have a small fraction of outliers, with HF-GP outliers tending towards larger MSEs. This gain in performance for VarP-GP (i.e., smaller MSE) is attributable primarily to the pooling of information from high-precision calculations over the parameter space, whereby the configuration of design points and event counts with varying precision is chosen to optimize the utilization of information from the training data (Eq.~\ref{eq:pair}). Notably, the MSE values for VarP-GP are also considerably lower than those for LF-GP, indicating that the optimal utilization of the event budget for emulation corresponds to a judiciously chosen balance of high and low-precision design point calculations.


\begin{figure*}[hbt!]
    \centering
    \includegraphics[width=0.495\linewidth]{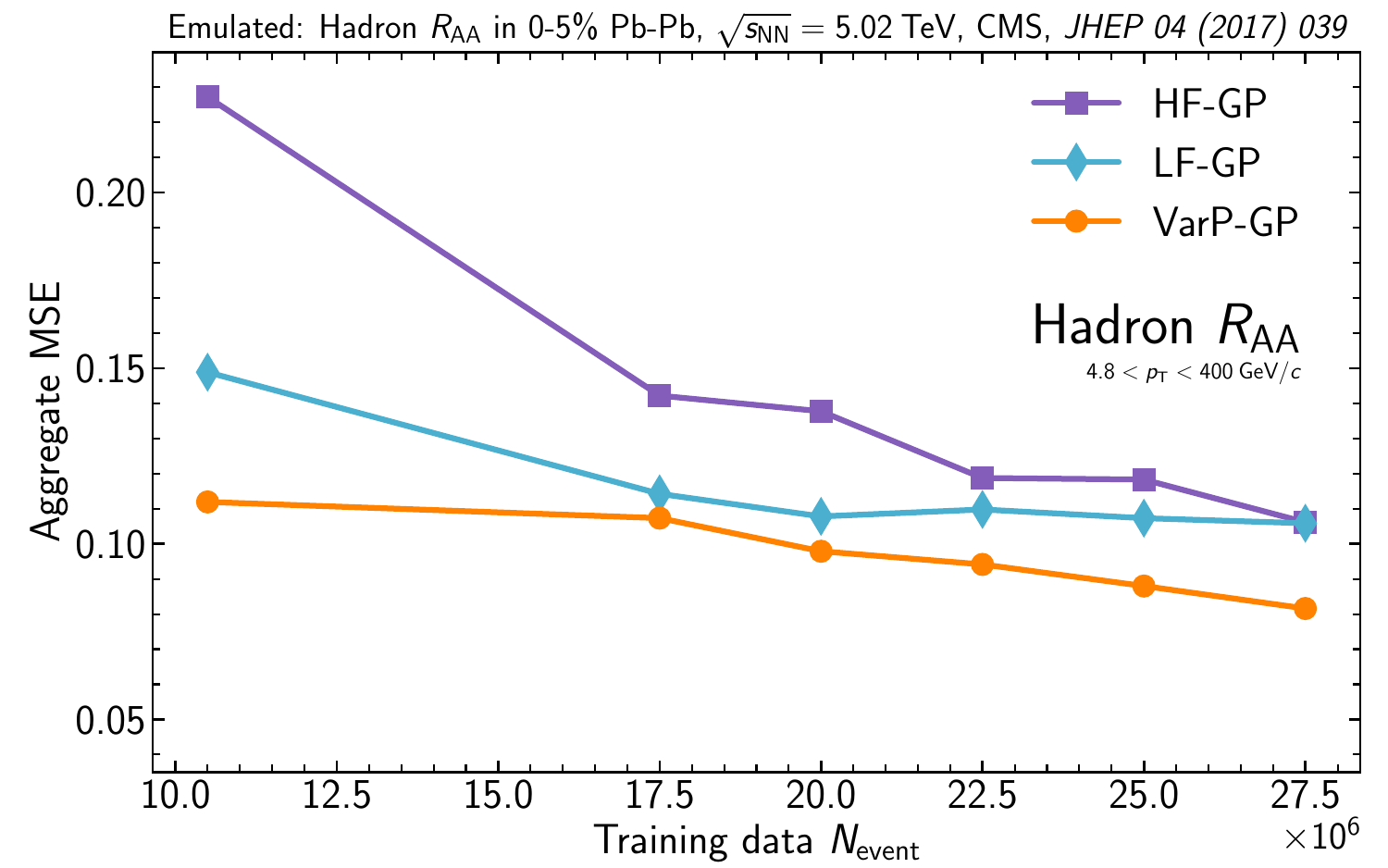}
    \includegraphics[width=0.495\linewidth]{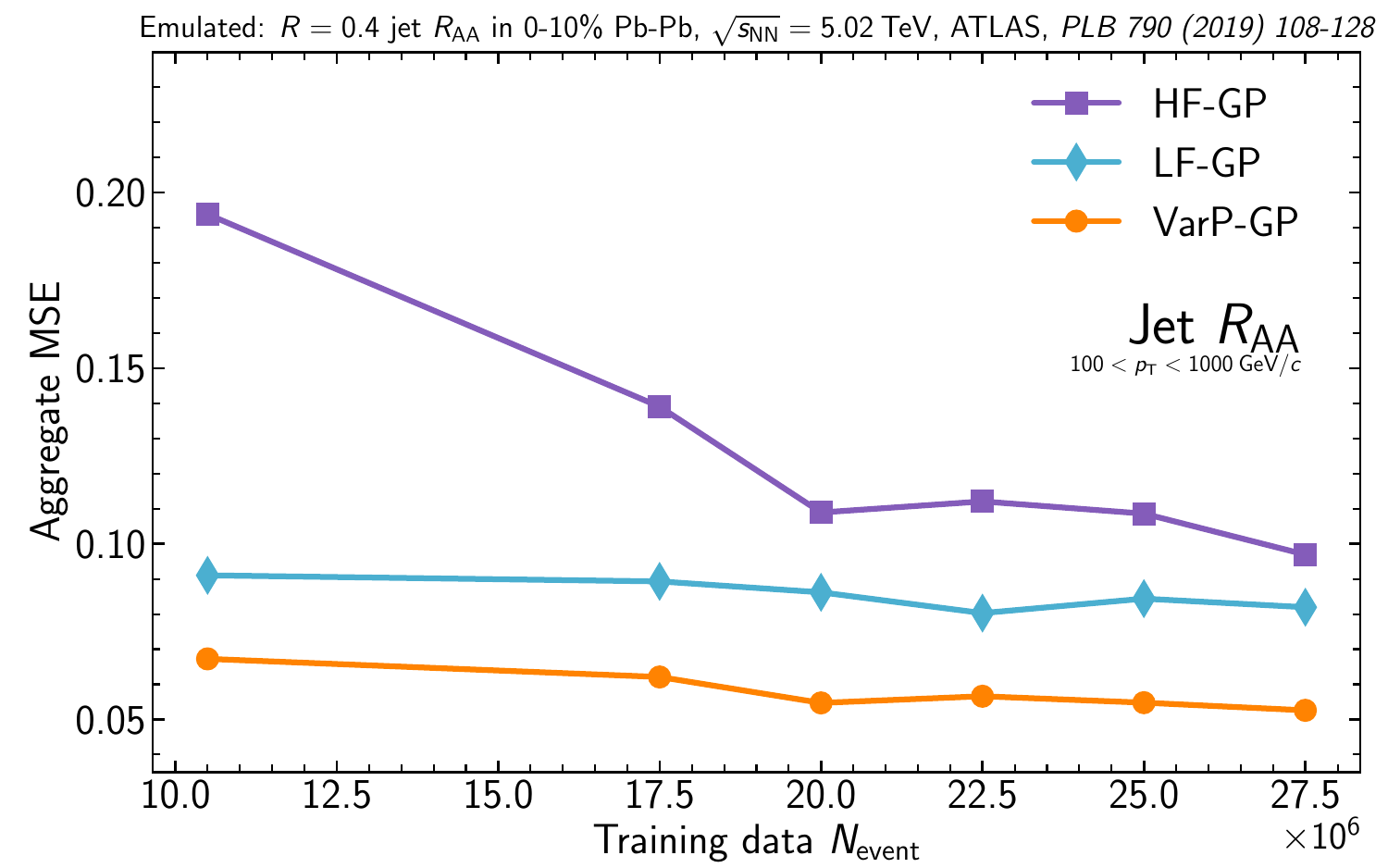}
    \caption{Aggregated \pT-differential MSE (sum of MSE over all \pT\ bins) from Fig.~\ref{fig:differentialEmulationPerformance}, as a function of the number of simulated events \Nevent\ used for emulator training. Left: hadron \RAA; right: jet \RAA. Orange: VarP-GP; blue: LF-GP; purple: HF-GP.}
    \label{fig:integreatedEmulateionPerformance}
\end{figure*}

Figure~\ref{fig:integreatedEmulateionPerformance} shows the aggregated \pT-differential MSE distributions from Fig.~\ref{fig:differentialEmulationPerformance}, corresponding to the sum of MSE over all \pT\ bins, as a function of \Nevent. The VarP-GP emulator again has noticeably smaller MSE than HF-GP and LF-GP, indicating improved emulation precision, with qualitatively similar trends for hadron and jet \RAA. For hadron \RAA, HF-GP requires about 60\% more training data to achieve the same level of emulation accuracy as VarP-GP. For jet \RAA, HF-GP cannot achieve the same level of emulation accuracy as VarP-GP over the full range of computing budget considered, by a wide margin. The performance of LF-GP is better than that of HF-GP, but is likewise inferior to that of VarP-GP. Fig.~\ref{fig:integreatedEmulateionPerformance} again demonstrates that cost-efficient emulation can be achieved by pooling information from high-precision calculations over the parameter space, via a carefully selected balance of high and low-precision design points. 

The HF-GP (for hadron and jet \RAA) and LF-GP (for hadron \RAA) emulators exhibit rapidly decreasing MSE with increasing \Nevent\ at low values of \Nevent, while VarP-GP shows less dependence on \Nevent\ for its MSE.
This feature corresponds to the well-established reduction in the rate of decrease for GP emulation error with increasing amount of training data~\cite{van2011information}. By pooling information from high and low-precision calculations, the VarP-GP emulator utilizes the event budget more efficiently, augmenting the effective amount of training data.

\subsection{Application of VarP-GP: parameter sensitivity}
\label{sect:SensitivityAnalysis}


As an initial application of the VarP-GP emulator we assess the sensitivity of physical observables to the changes in model parameters, based on the Bayesian Inference analysis for jet quenching carried out in Ref.~\cite{JETSCAPE:2024cqe}. Sensitivity assessments are employed broadly in Bayesian uncertainty quantification for complex physical systems \cite{smith2024uncertainty}. However, they can be computationally expensive due to the large number of calculations required; in this case such an assessment would be intractable on achievable computing resources if the full physics model were used. We therefore substitute direct computation of the full physics model with emulation, at much lower computational cost. We compare the performance of VarP-GP and HF-GP as a qualitative cross-check, in place of the inaccessible ground truth sensitivity.

While a sensitivity analysis has been carried out for bulk QGP observables~\cite{JETSCAPE:2020mzn, Nijs:2020roc, Parkkila:2021yha, liyanage2022efficient}, it has not yet been applied to jet quenching data.

The sensitivity analysis utilizes the variance-based Sobol' indices~\cite{sobol1990sensitivity,sobol1993sensitivity} which were used in previous QGP studies~\cite{liyanage2022efficient}. The normalized total-effect Sobol' index~\cite{JACQUES20061126,saltelli2010variance} is defined as

\begin{align}
\begin{split}
    & \STi = \frac{\mathbb{E}_{\XbfTildei}[\text{Var}_{\Xsubi}\{f(\mathbf{X})|\XbfTildei\}]}
    {\text{Var}_{\Xbf}\{f(\mathbf{X})\}},\\
    &\STinorm = \frac{\STi}{\sum_{l=1}^d{\STi}}, \hspace{0.75cm} l = 1, \cdots, d.
    \label{eq:sobelTotalIndex}
\end{split}
\end{align}

\noindent
Here, $\Xbf=(X_1,\cdots,X_d)$ is a vector of $d$ model parameters; $X_{l}$ is an independent uniform random variable for the $l$-th model parameter over the range considered; and $f(\mathbf{x})$ is the value of a physical observable (e.g. \RAA(\pT)) for model parameter vector $\mathbf{x}$. The notation \XbfTildei\ indicates the set of all parameters in \Xbf\ except for \Xsubi. The expected value $\mathbb{E}_{\XbfTildei}[\text{Var}_{\Xsubi}\{f(\mathbf{X})|\XbfTildei\}]$
is the expected variance of the observable when all parameters except \Xsubi\ are held fixed.
The Sobol' index \STi\ thus measures the relative contribution of parameter \Xsubi\ to the total variance of $f$ (namely, $\text{Var}_{\Xbf}\{f(\mathbf{X})\}$), including the effects of the covariance of \Xsubi\ with other parameters $j \neq l$.

It can be shown that $\sum_{l=1}^d \STi$, the sum of the total-effect Sobol' indices, exceeds unity. The normalized index \STinorm\ is however bounded by $[0, 1]$, with the parameters that are most sensitive to the values of the physical observables exhibiting the largest values. The normalized index \STinorm\ therefore gauges the relative contribution of \Xsubi\ to remaining parameters. The physics model in Ref.~\cite{JETSCAPE:2024cqe}, which is described in Sect.~\ref{sect:physicsModelBayesianAnalysis}, has $d=6$ parameters. The case of equal sensitivity to all parameters corresponds to $\STinorm=1/6$.

This study employs two approaches for sensitivity analyses based on Sobol' indices. The first, referred to as the global analysis, evaluates the sensitivity measure over the entire (global) design space. The second, referred to as the local analysis, evaluates the sensitivity measure over a restricted (local) region of interest. The local region can be selected near the maximum a-posteriori (MAP) parameters from a Bayesian analysis, to assess model behavior near the most plausible physical parameters. The global and local analyses are complementary, providing insight into model sensitivity in different regions of  the parameter space.

The sensitivity analysis explores the following model parameters~\cite{JETSCAPE:2024cqe}:
\begin{itemize}
    \item \alphas\ (denoted \alphasfix\ in Eq.~4 in Ref.~\cite{JETSCAPE:2024cqe}), the coupling at the soft scale, with range $0.1 \leq \alphas \leq 0.5$.
    \item \qswitch, the transition scale between the higher (MATTER) and lower (LBT) virtuality stages of the simulation, with range $1~\mathrm{GeV}\leq{\qswitch}\leq10~\mathrm{GeV}$.
    \item \tstart, the start time of jet modification, with range $0\leq \tstart\leq 1.5$~fm/$c$.
    \item Parameters $c_1 , c_2, c_3$, which control the modification of \qhat\ with increasing virtuality, with range $-5 \leq \log(c_{1,2}) \leq \log(10)$ and $-3 \leq \log(c_3) \leq \log(100)$.
\end{itemize}
The global analysis covers the parameter ranges defined above, while the local analysis is restricted to the 1-99\% range of the posterior distributions found in Ref.~\cite{JETSCAPE:2024cqe}.

The data employed are the same as those used in Sect.~\ref{sect:EmulatorPerformance}, namely the hadron and jet \RAA\ for central \PbPb\ collisions at $\sqrtsNN=5.02$ TeV~\cite{CMS:2016xef,ATLAS:2018gwx}. Values of \RAA\ are calculated by the emulator using the data binning, and are then grouped to explore the \pT\ evolution. Hadron \RAA\ is grouped into four bins, while jet \RAA\ is grouped into three bins. The observed inclusive hadron population is dominated by leading hadrons of a jet that carry large jet momentum fraction $z$. The \pT\ ranges of the hadron and jet data are selected to account for this effect, such that the hadron and jet distributions originate from a similar population of hard-scattered partons.

The sensitivity indices are then computed using the Python \texttt{SAlib} library \citep{Herman2017,Iwanaga2022}.
For the global analysis, $5,000\times(2\times d+2)=70,000$ sample points are generated over the full parameter space using the standard Saltelli method \cite{saltelli2010variance}, which is the \texttt{SAlib} library default option. For each \pT\ bin, predictions are generated at each sample point by the VarP-GP and HF-GP emulators. These predictions are then input to the \texttt{sobol.analyze} function from the same package, to compute the total-effect indices $\{\STi\}_{l=1}^d$ and their associated uncertainties for each \pT\ bin. Finally, the computed indices are aggregated across \pT\ bins, their systematic uncertainties are evaluated using the law of total variance \cite{billingsley2013convergence}, and they are normalized following Eq.~\ref{eq:sobelTotalIndex}. The same procedure is used for the local analysis, with the sample points now taken from the 1-99\% range of the posterior distributions found in Ref.~\cite{JETSCAPE:2024cqe}.

\begin{figure*}[hbt!]
    \centering
    \includegraphics[width=0.7\linewidth]{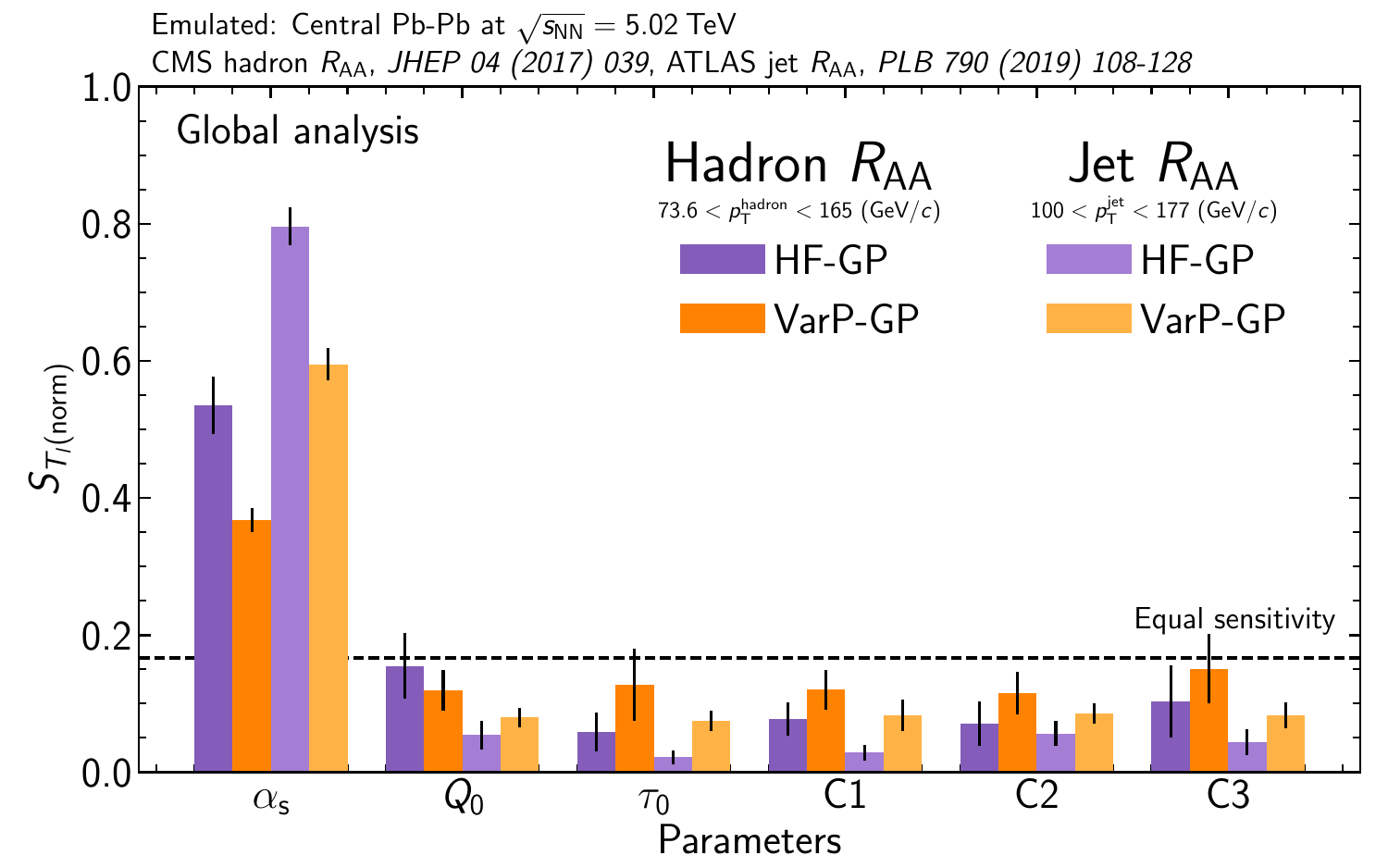}
    \caption{Values of \STinorm\ using HF-GP and VarP-GP for the global analysis over the full design space, for hadron and jet \RAA\ in \pT\ ranges with comparable partonic kinematics. Systematic uncertainties are shown as black vertical lines. The dashed horizontal line shows the value of \STinorm\ for equal sensitivity (i.e. 1/6).}
    \label{fig:hadronVsJetSensitivity}
\end{figure*}

Figure~\ref{fig:hadronVsJetSensitivity} shows the global sensitivity analysis over the full design space for the VarP-GP and HF-GP emulators, for hadron and jet \RAA\ in \pT\ ranges with comparable partonic kinematics. Both emulators are trained using the largest target event counts at all design points. The emulators exhibit qualitatively similar behavior for both observables. The parameter \alphas\ is the most sensitive parameter by a wide margin in all cases, with the remaining sensitivity distributed approximately uniformly among the other parameters. This finding corresponds to the physical intuition that hadron and jet \RAA\ at high \pT\ are sensitive primarily to overall jet energy loss, which is driven primarily by the coupling strength \alphas~\cite{JETSCAPE:2024cqe}, rather than finer details of jet energy redistribution due to quenching. 

This figure also shows that, for both emulators, sensitivity to the value of \alphas\ is higher for jet than for hadron \RAA. While it is of physics interest to understand this effect, its origin may be complex and elucidating it requires a separate study that is beyond the scope of this analysis.

\begin{figure*}[hbt!]
    \centering
    \includegraphics[width=0.495\linewidth]{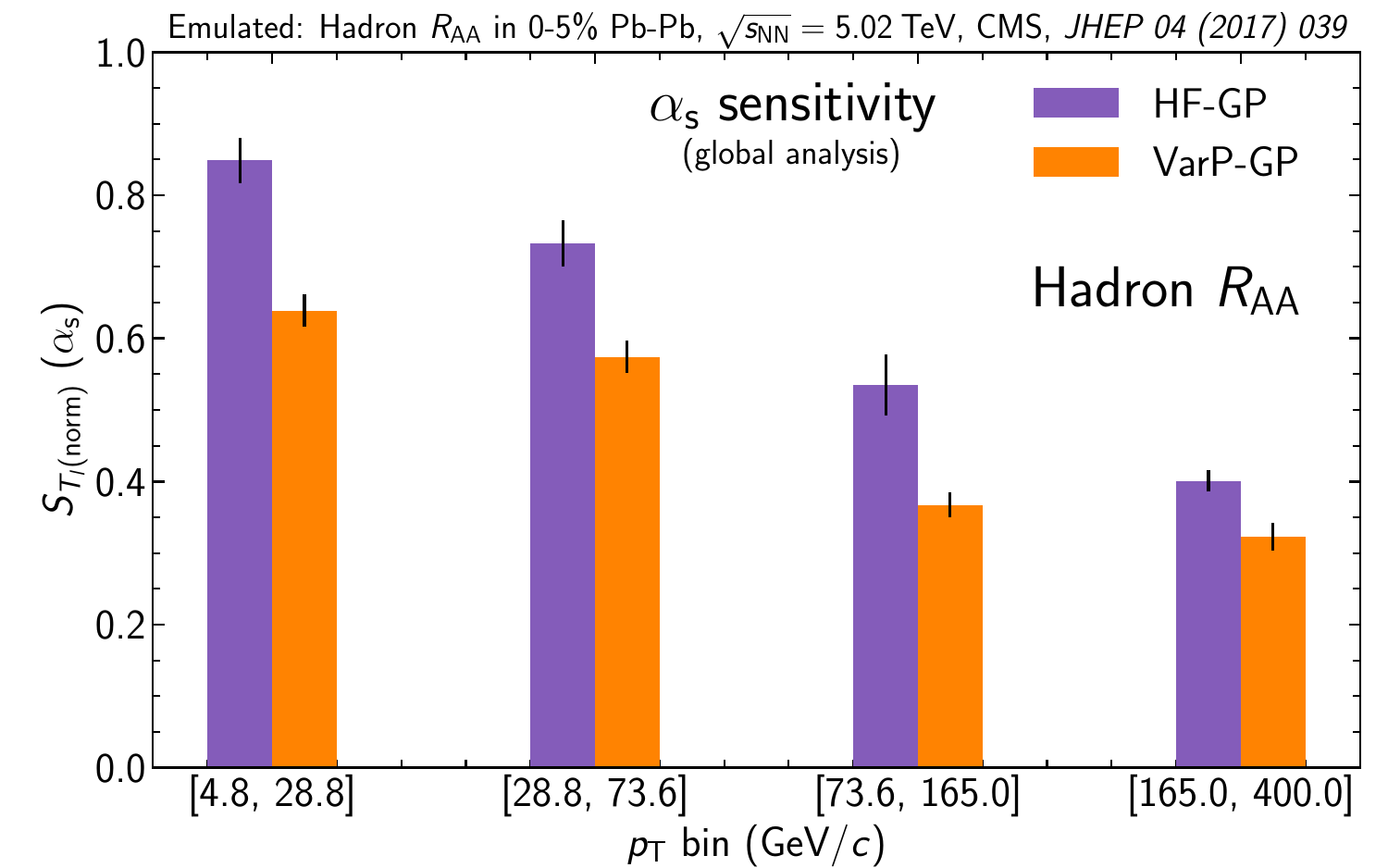}
    \includegraphics[width=0.495\linewidth]{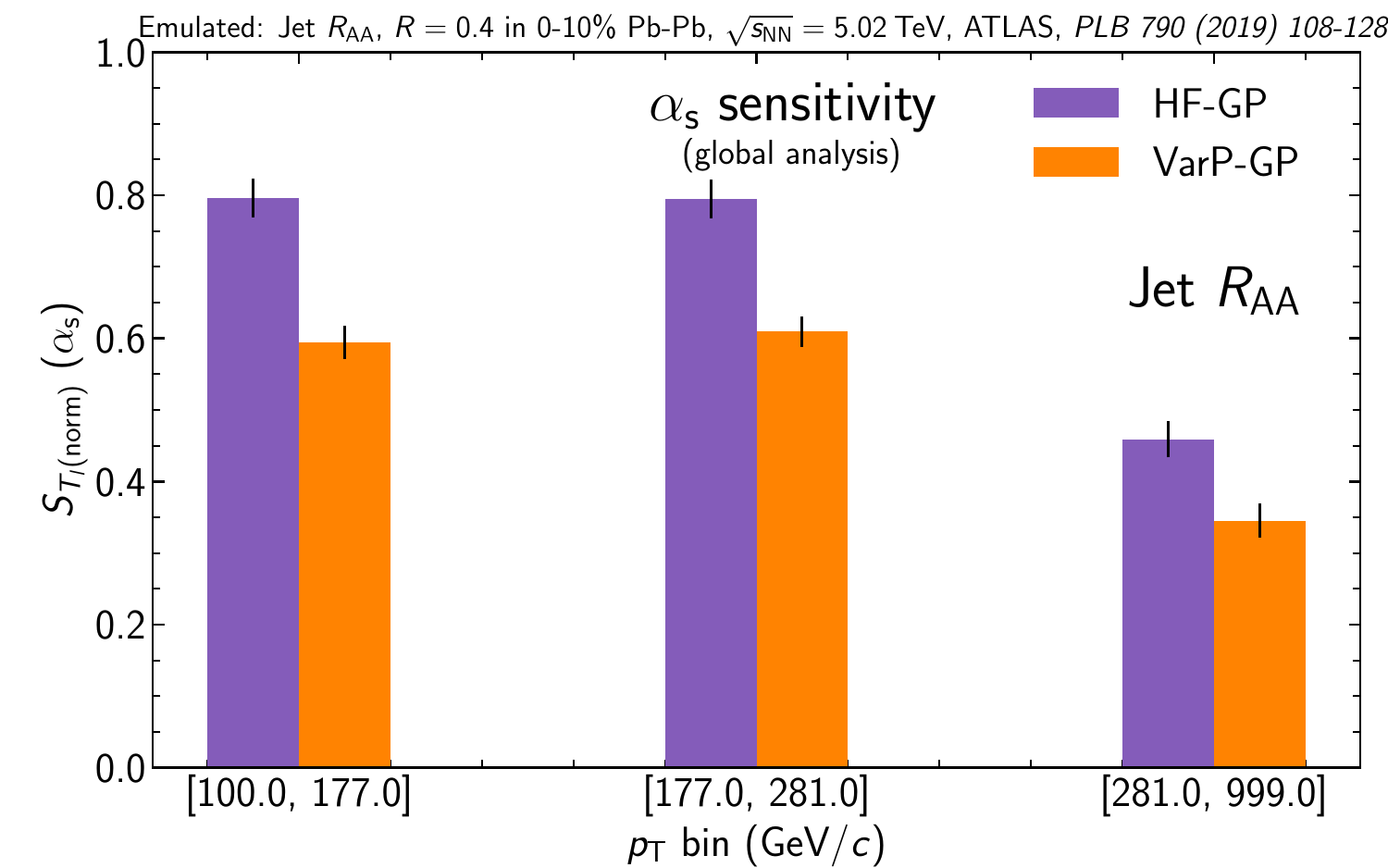}
    \caption{Values of \STinorm\ for \alphas\ the global analysis over the full design space, for hadron (left) and jet (right) \RAA\  in selected \pT\ ranges. Systematic uncertainties are shown as black vertical lines.
    }
    \label{fig:alphasSensitivity}
\end{figure*}

Figure~\ref{fig:alphasSensitivity} compares the \alphas\ sensitivity of VarP-GP and HF-GP in selected \pT\ bins.
While the overall sensitivity decreases with increasing \pT\ for both hadrons and jets, the greater sensitivity of HF-GP than VarP-GP persists.
The \pT-dependence of the diffusion of sensitivity to other model parameters suggests additional physical processes play a larger role at high \pT. Detailed model studies are required to elucidate such effects, which however are beyond the scope of the current analysis.

\begin{figure*}[hbt!]
    \centering
    \includegraphics[width=0.495\linewidth]{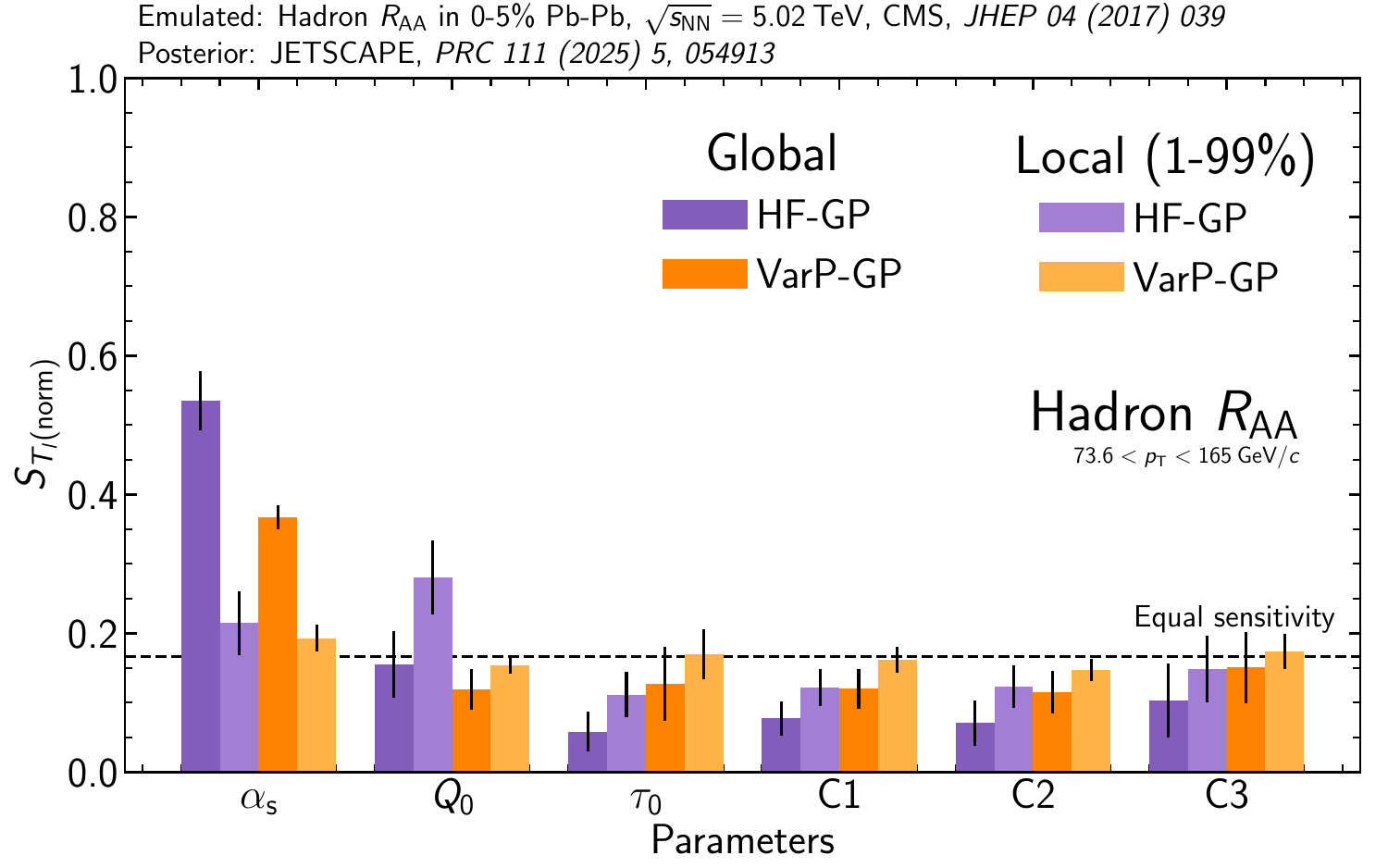}
    \includegraphics[width=0.495\linewidth]{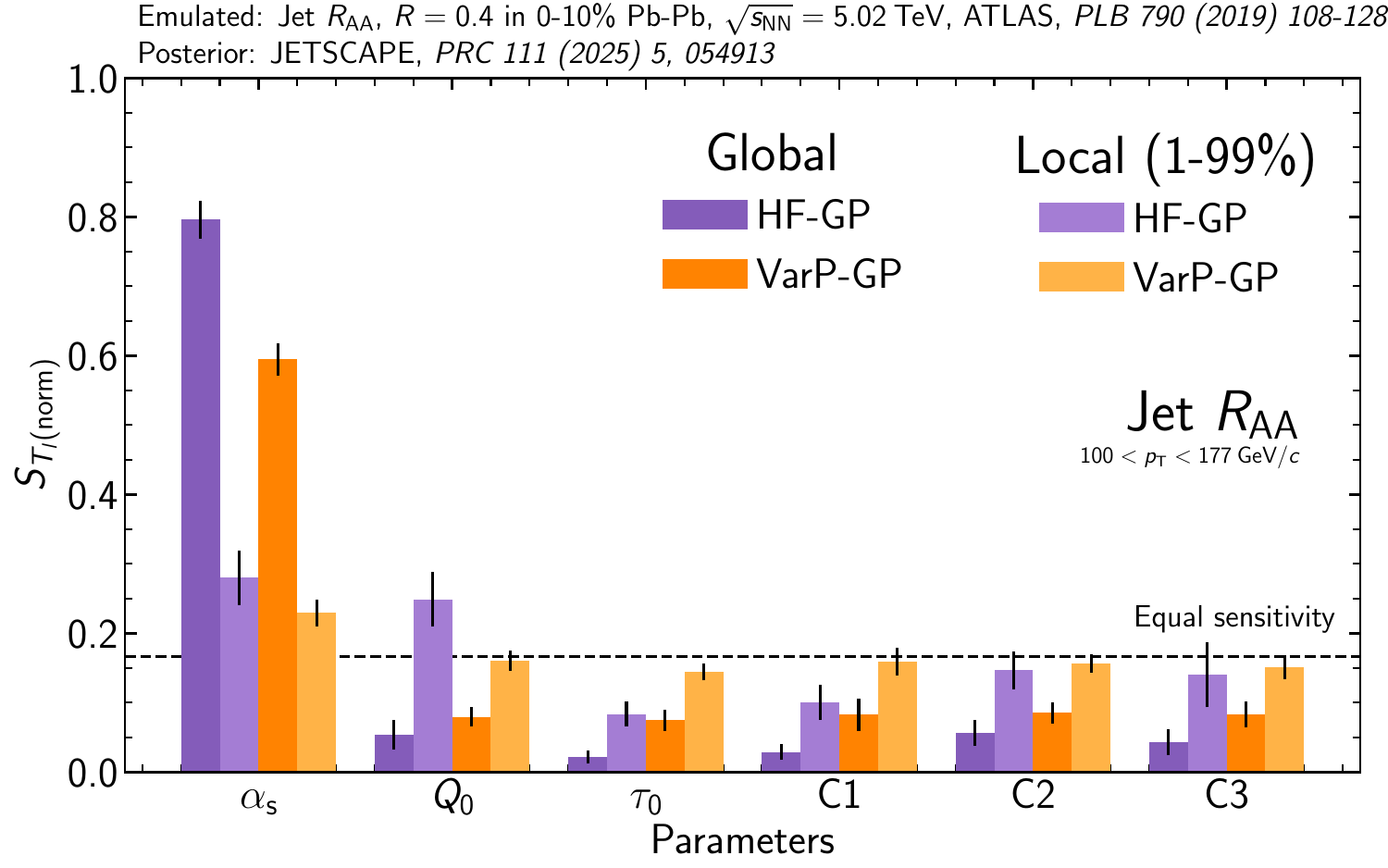}
    \caption{Values of \STinorm\ from VarP-GP and HF-GP for the global and local sensitivity analyses, for hadron (left) and jet (right) \RAA in \pT\ ranges with comparable partonic kinematics. Systematic uncertainties are shown as black vertical lines. The dashed horizontal line shows the value of \STinorm\ for equal sensitivity (i.e., 1/6).
    }
    \label{fig:globalVsLocalSensitivity}
\end{figure*}

Figure~\ref{fig:globalVsLocalSensitivity} compares the results of the global and local sensitivity analyses. For both HF-GP and VarP-GP the local-analysis sensitivity, in which model parameters are constrained near the MAP values, is markedly more balanced than that of the global analysis. Balanced sensitivity for all parameter values close to the MAP parameters suggests that several processes, which are governed by different parameters, have similar influence near the region of physical interest. There may be additional differences between the emulators in other aspects of parameter sensitivity, which are secondary in magnitude to the main effect of an overall greater balance in sensitivity for the local analysis, and their exploration is beyond the scope of this work.

The difference between global and local analyses is the exclusion from the latter of less--likely outlier parameter sets. It is evident from the comparison of Figs.~\ref{fig:hadronVsJetSensitivity}, \ref{fig:alphasSensitivity}, and \ref{fig:globalVsLocalSensitivity} that HF-GP is more susceptible than VarP-GP to pulls from outlier parameter sets. This may arise because HF-GP is less aware of the overall contours of the parameter space, and may therefore give additional emphasis to corners of parameter space that are far from the MAP solution.

This sensitivity study provides additional evidence that VarP-GP provides a more robust approach to inference computations than HF-GP, in addition to the marked improvement in computational efficiency demonstrated in Sect.~\ref{sect:EmulatorPerformance}. The full parameter dependence and \pT\ evolution of the global and local sensitivity analyses are included in App.~\ref{app:SensitivityGallery}, to enable additional in-depth physics model investigations of the general trends reported here.

\section{Summary}
\label{sect:summary}

This manuscript presents VarP-GP, a cost-efficient Bayesian emulator for expensive computational models with variable statistical precision. The VarP-GP model utilizes statistical precision as a measure of simulation fidelity, efficiently allocating a given computational budget by optimizing the distribution of high and low-fidelity calculations over the parameter design space. The algorithm has two components: a space-filling experimental design to maximize available information, and an uncertainty-aware Gaussian process emulator to exploit the precision metric. 

The performance of VarP-GP is explored using JETSCAPE calculations of jet quenching in the quark-gluon plasma. Comparison is made to a conventional GP emulators, LF-GP and HF-GP, which distribute the computing budget uniformly to larger or smaller number of design points. The VarP-GP has better performance than either conventional emulator in terms of both computational precision achieved for fixed computing budget, and computing budget required for a given precision. This indicates that exploration of the entire design space with mixed precision levels is the optimal utilization of a given computing resource budget.

As an initial physics application of VarP-GP, the computationally-expensive exploration of the model parameter sensitivity based on physical observables is carried out for a recent Bayesian Inference calibration of jet quenching by JETSCAPE. This study likewise shows superior performance of VarP-GP compared to HF-GP due to the reduced influence of outlier design points that are far from the optimum found by the calibration.

The VarP-GP emulator provides a new approach to Bayesian emulation for computationally expensive calculations. Given the promising results presented here, we are exploring several directions for future work. First, we aim to combine the VarP-GP with other cost-efficient algorithms such as transfer learning \cite{liyanage2022efficient,wang2024local}, to perform novel Bayesian Inference analyses that would otherwise be prohibitively expensive, notably the interpretation of collider data measuring the structure and dynamics of the quark-gluon plasma. Second, we are developing new active learning algorithms~\cite{song2023ace} that leverage the fitted VarP-GP model to dynamically select new design points and their corresponding number of events for cost-efficient emulator training.

 
\section{Acknowledgments}

We thank the JETSCAPE Collaboration for providing the simulation data used in this work. This work was supported in part by the National Science Foundation (NSF) within the framework of the JETSCAPE collaboration, under grant number OAC-2004571 (CSSI:X-SCAPE). It was also supported in part by the US Department of Energy, Office of Science, Office of Nuclear Physics under grant number DE-AC02-05CH11231, and by the NSF Division of Mathematical Sciences under grant number 2316012 (Y.J., S.M.).

\appendix

\section{Derivation of VarP-GP likelihood function}
\label{app:deriv}

\noindent In what follows, we derive the likelihood expression in Eq.~\eqref{eq:lkhd} for estimating hyperparameters of the VarP-GP. Adopt the earlier notation for the simulated observables $\mathbf{y}$, its precision matrix $\mathbf{P}$, and the corresponding variance vector $\mathbf{s}^2$. When both $f(\cdot)$ and $\log s^2(\cdot)$ follow independent GP priors, one can show using the multivariate normal property of a GP that $\mathbf{y}$ and $\log\mathbf{s}^2$ follow the multivariate normal distributions:
\begin{equation}
    \mathbf{y}\sim \mathcal{N}(\mu_f\mathbf{1}_n,\mathbf{K}_f+\mathbf{P}^{-1}) \quad \text{and} \quad \log\mathbf{s}^2\sim \mathcal{N}(\mu_s\mathbf{1}_n,\mathbf{K}_s).
\end{equation}
As before, $\mathbf{K}_f = [k_f(\mathbf{x}_i,\mathbf{x}_j)]_{i,j=1}^n$ and $\mathbf{K}_s = [k_s(\mathbf{x}_i,\mathbf{x}_j)]_{i,j=1}^n$ are the data covariance matrices for $f$ and $\log s^2$, respectively.

Using these multivariate normal distributions, the joint likelihood of the set of hyperparameters $\boldsymbol{\Theta} = \{\mu_f,\mu_s,\sigma^2_f,\sigma^2_s,\boldsymbol{\theta}_f,\boldsymbol{\theta}_s\}$ can be written as:
\small
\begin{align}
\begin{split}
\mathcal{L}(\boldsymbol{\Theta}) &= \mathcal{L}(\mathbf{y}|\mathbf{s}^2)\times \mathcal{L}(\mathbf{s}^2)\\
&=(2\pi)^{-\frac{n}{2}}|\mathbf{K}_f+\mathbf{P}^{-1}|^{-\frac{1}{2}}\\
& \quad \times \exp\left\{-\frac{1}{2}(\mathbf{y}-\mu_f\mathbf{1}_n)^T(\mathbf{K}_f+\mathbf{P}^{-1})^{-1}(\mathbf{y}-\mu_f\mathbf{1}_n)\right\}\\
& \quad \times (2\pi)^{-\frac{n}{2}}|\mathbf{K}_s|^{-\frac{1}{2}} \\
& \quad \times \exp\left\{-\frac{1}{2}(\log\mathbf{s}^2-\mu_s\mathbf{1}_n)^T\mathbf{K}_s^{-1}(\log\mathbf{s}^2-\mu_s\mathbf{1}_n)\right\}.
\end{split}
\end{align}
\normalsize
\noindent From this, the log-likelihood takes the form:
\small
\begin{align}
\begin{split}
l(\boldsymbol{\Theta}) &= -\frac{1}{2}\log|\mathbf{K}_f+\mathbf{P}^{-1}|\\
& \quad -\frac{1}{2}(\mathbf{y}-\mu_f\mathbf{1}_n)^T(\mathbf{K}_f+\mathbf{P}^{-1})^{-1}(\mathbf{y}-\mu_f\mathbf{1}_n) \\
&\quad \quad - \frac{1}{2}\log|\mathbf{K}_s| - \frac{1}{2}(\log\mathbf{s}^2)^T\mathbf{K}_s^{-1}(\log\mathbf{s}^2) + C,
\end{split}
\end{align}
\normalsize
where $C$ captures all constants not depending on $\boldsymbol{\Theta}$ (and thus are not important for the optimization of $l(\boldsymbol{\Theta})$).

With the model hyperparameters $\boldsymbol{\Theta}$ optimized via the maximization of $l(\boldsymbol{\Theta})$, the posterior predictive equations in Eqs.~\eqref{eq:gphet1}-\eqref{eq:gphet3} follow from the GP predictive equations in Eqs.~\eqref{eq:gpconst1}-\eqref{eq:gpconst2} and the conditional independence of the stochastic processes $[f(\cdot)|\mathbf{y},\mathbf{p}]$ and $[\log s^2(\cdot)|\mathbf{p}]$.

\section{Full gallery of hadron and jet \texorpdfstring{$\RAA{}$}{RAA} sensitivity}
\label{app:SensitivityGallery}

\begin{figure*}[hbt!]
    \centering
    \includegraphics[width=0.5\linewidth]{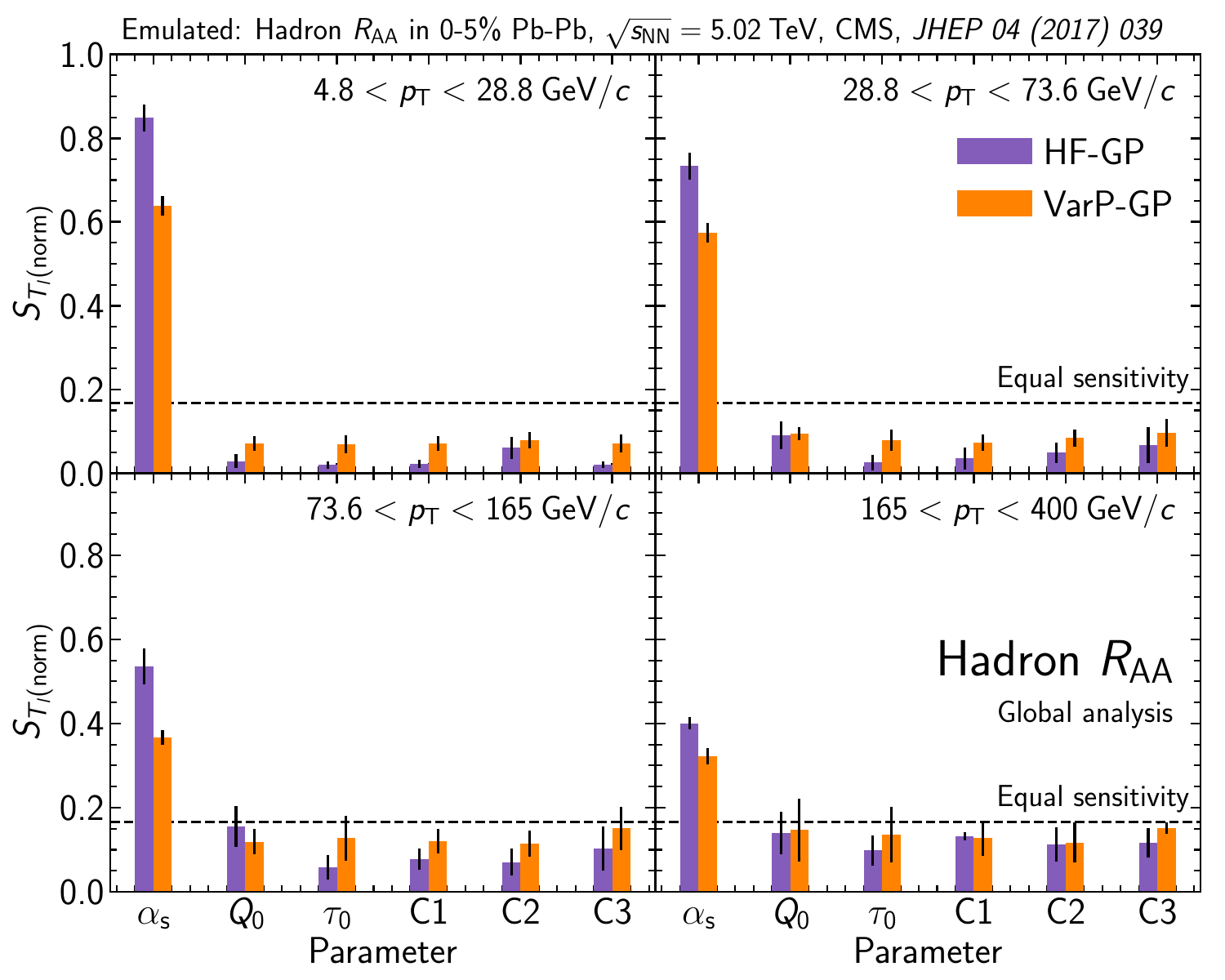}
    \caption{Values of \STinorm\ for hadron $\RAA{}$ in selected $p_{\text{T}}$ bins using full design space. The HF-GP is shown in purple, while the VarP-GP is shown in orange. Equal sensitivity is shown with the dashed black horizontal line. Systematic uncertainties are shown as black vertical lines.}
    \label{fig:sensitivityHadronGlobalPtDiff}
\end{figure*}

\begin{figure*}[hbt!]
    \centering
    \includegraphics[width=0.5\linewidth]{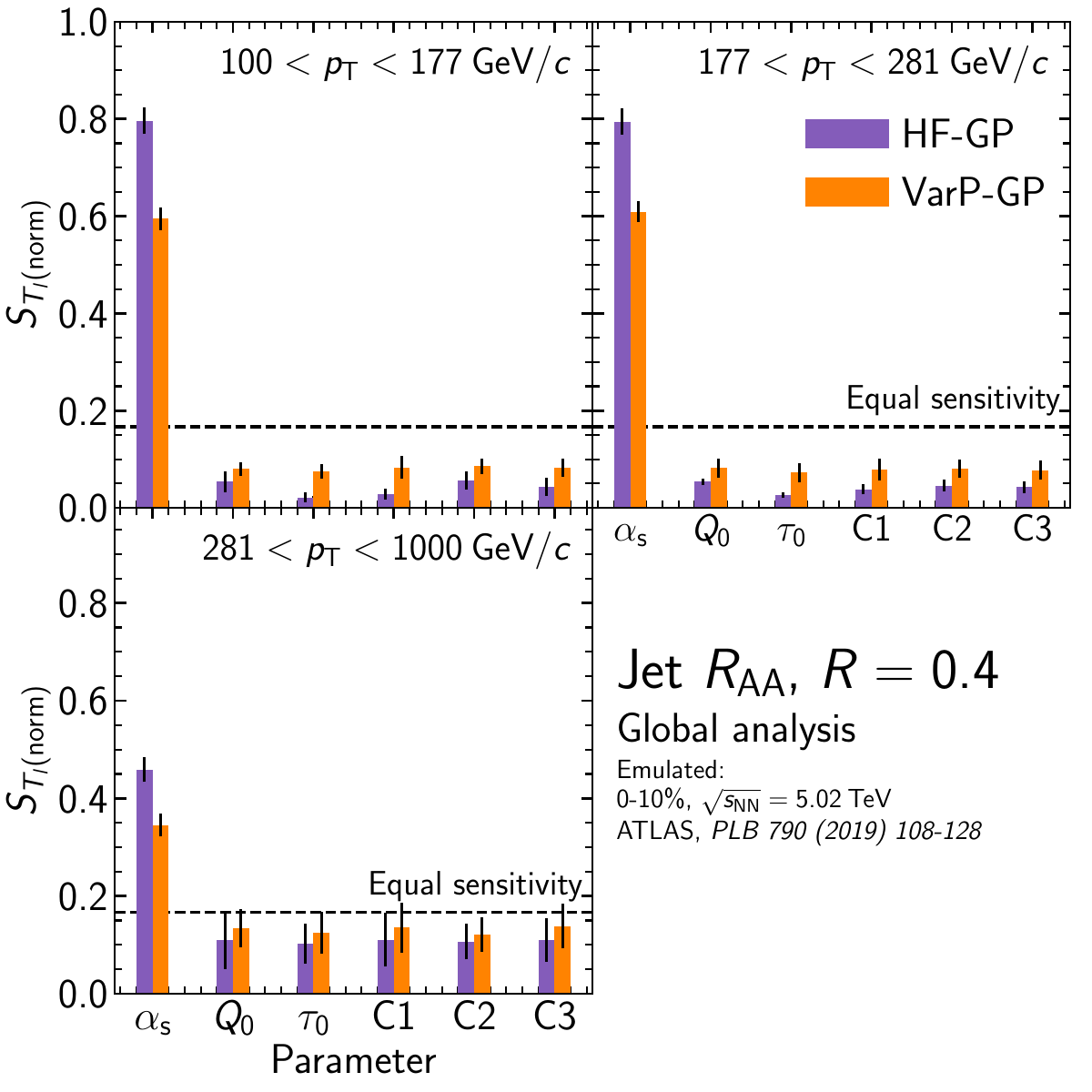}
    \caption{Values of \STinorm\ for jet $\RAA{}$ in selected $p_{\text{T}}$ bins using full design space. The high fidelity GP is shown in purple, while the VarP-GP is shown in orange. Equal sensitivity is shown with the dashed black horizontal line. Systematic uncertainties are shown as black vertical lines.}
    \label{fig:sensitivityJetGlobalPtDiff}
\end{figure*}

\begin{figure*}[hbt!]
    \centering
    \includegraphics[width=0.5\linewidth]{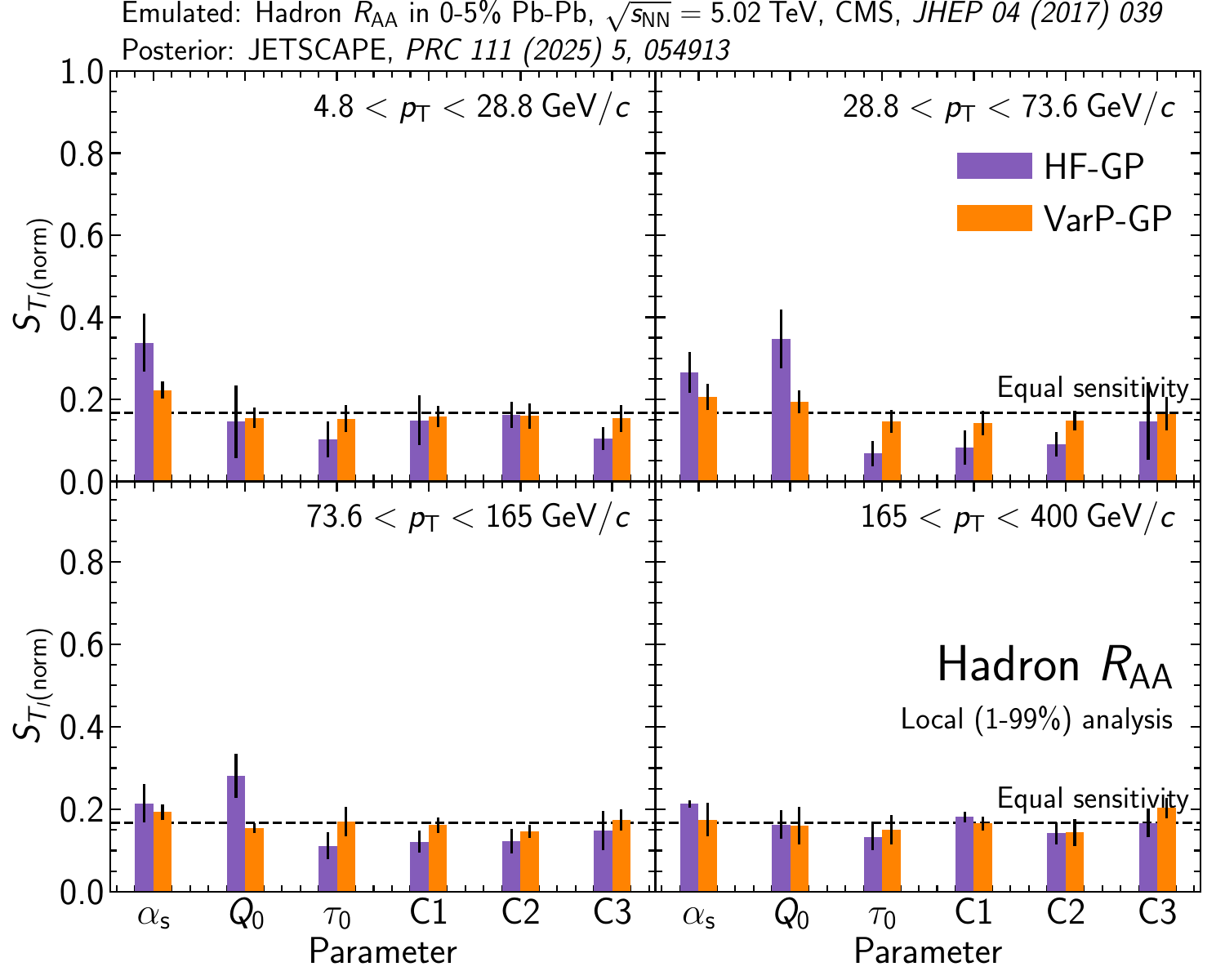}
    \caption{Values of \STinorm\ for hadron $\RAA{}$ in selected $p_{\text{T}}$ bins using the 1-99\% posterior. The high fidelity GP is shown in purple, while the VarP-GP is shown in orange. Equal sensitivity is shown with the dashed black horizontal line. Systematic uncertainties are shown as black vertical lines.}
    \label{fig:sensitivityHadronLocalPtDiff}
\end{figure*}

\begin{figure*}[hbt!]
    \centering
    \includegraphics[width=0.5\linewidth]{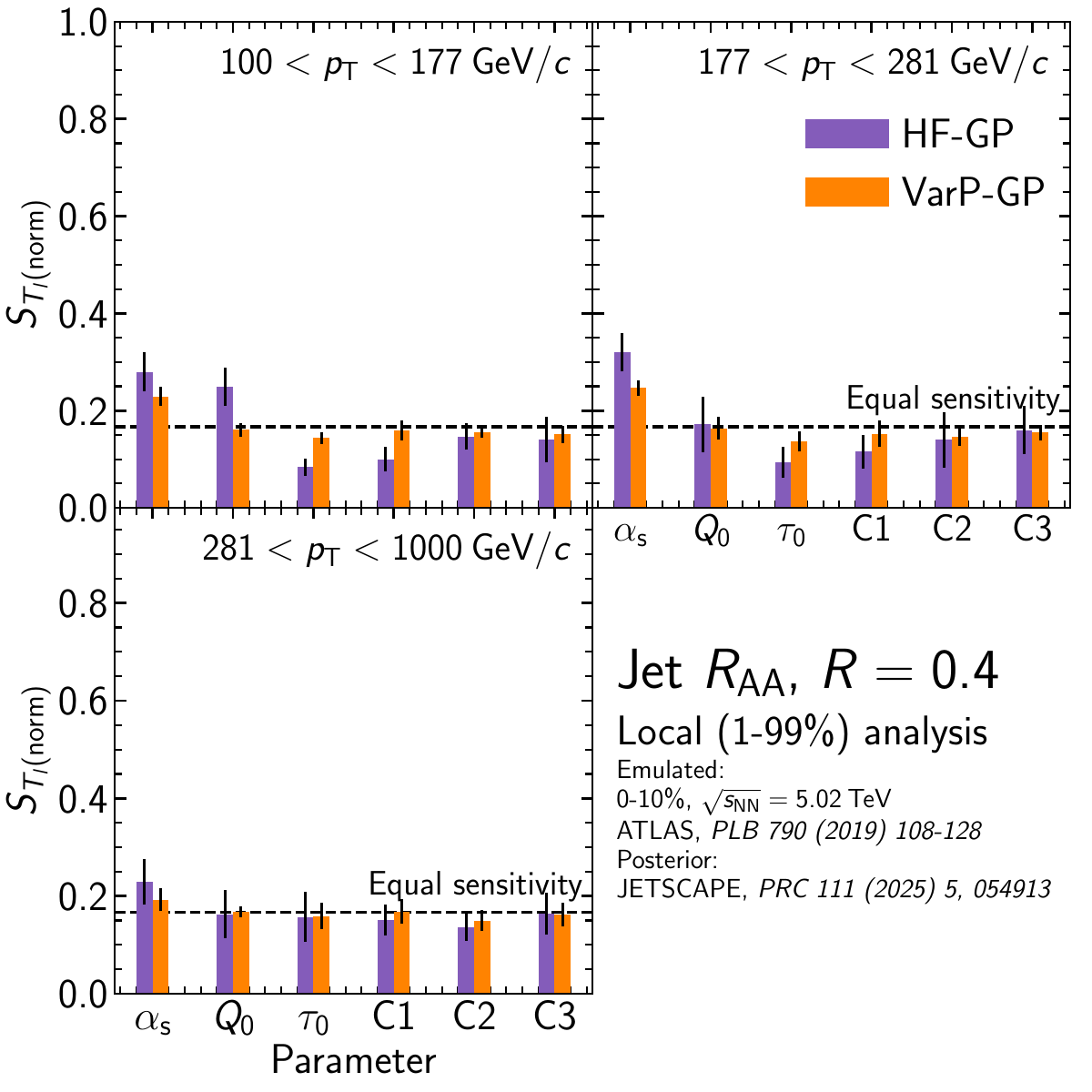}
    \caption{Values of \STinorm\ for jet $\RAA{}$ in selected $p_{\text{T}}$ bins using the 1-99\% posterior. The high fidelity GP is shown in purple, while the VarP-GP is shown in orange. Equal sensitivity is shown with the dashed black horizontal line. Systematic uncertainties are shown as black vertical lines.}
    \label{fig:sensitivityJetLocalPtDiff}
\end{figure*}

The observations presented above are based on a curated selection of the parameter sensitivity studies. This appendix presents the full $\pT{}$-dependence of the sensitivity (Sect.~\ref{sect:SensitivityAnalysis}) for both the local and global analyses, as input to future model studies which investigate the physical mechanisms underlying the parameter sensitivity. The global analysis is shown in Fig.~\ref{fig:sensitivityHadronGlobalPtDiff} for the hadron $\RAA{}$ and in Fig.~\ref{fig:sensitivityJetGlobalPtDiff} for the jet $\RAA{}$. The local 1-99\% posterior analysis is shown in Fig.~\ref{fig:sensitivityHadronLocalPtDiff} for the hadron $\RAA{}$ and in Fig.~\ref{fig:sensitivityJetLocalPtDiff} for the jet $\RAA{}$.

\bibliographystyle{utphys}   
\bibliography{bibliography}

\end{document}